# An extended Kolmogorov–Avrami–Ishibashi (EKAI) model to simulate dynamic characteristics of polycrystalline-ferroelectric-gate field-effect transistors

Shigeki Sakai[1,2,*] and Mitsue Takahashi[1,§]

[1]*National Institute of Advanced Industrial Science and Technology, 1-1-1, Umezono, Tsukuba, Ibaraki 305-8568, Japan*

[2]*Research Center for Neuromorphic AI Hardware, Kyushu Institute of Technology, Fukuoka, Japan*

Abstract

An extended Kolmogorov–Avrami–Ishibashi (EKAI) model is proposed, which represents dynamic characteristics in polycrystalline ferroelectric films consisting of grains under time-dependent electric fields. The original Kolmogorov–Avrami–Ishibashi (KAI) model described time-varying polarization reversal in a single-crystal film under a constant electric field. The field and the polarization were directed along the film normal. The polarization reversal dynamics were represented by a time-evolution function, $c(t)$, [Eq. (2) of this paper] regarding the volume fractions of the downward- and upward-polarization domains under a constant electric field. In the EKAI model, a grain in a polycrystalline ferroelectric film is indexed by $l$ with an angle $\theta_l$ which is the angle between the spontaneous polarization $P_s$ and the film normal. The EKAI model first assumes KAI-like polarization variations under a constant field inside the grain whose angle is tilted by $\theta_l$. The second assumption is concerned with polarization variation under time-varying electric field. During a time period $\Delta t$ from $t_{now}$ to $t_{next}(= t_{now} + \Delta t)$, the volume fractions of the downward- and upward-polarization domains changes according to the function $c(t)$ under an electric field at $t_{now}$. By combining the derived polarization and an electrostatic potential equation across a gate stacked structure at $t_{next}$, an electric field at $t_{next}$ is derived. Since the electric field and the volume fractions of the downward and upward domains are known at $t_{next}$, the calculation at the next $\Delta t$ step is possible. This procedure is repeated, and the EKAI model can simulate time-varying polarization reversal phenomena. The EKAI is applied for simulating quasi-static and dynamic characteristics of ferroelectric gate field-effect transistors (FeFETs) of $(Sr_{1-x}Ca_x)Bi_2Ta_2O_9$ (SBT). The SBT FeFETs are suitable for the model verification because they are reproducible with high endurance performance due to fairly small charge injection and trapping in the FeFET gate stacks. Consequent good agreement of the numerical results with the experimental indicates that the EKAI model is appropriate to simulate quasi-static and dynamic characteristics of FeFETs with polycrystalline ferroelectric films under swept and pulsed gate-voltages.

*shigeki.sakai@aist.go.jp

§mitsue-takahashi@aist.go.jp

## I. INTRODUCTION

Ferroelectric-gate field-effect transistors (FeFETs) have attracted attention not only by their potential functionality [1,2] such as compact, non-volatile, and non-destructive-read memory cells but also by recent rich accumulations of experimental data. FeFETs have been characterized by various measurements. Quasi-static characteristics of drain current ($I_d$) *versus* gate voltage ($V_g$) are measured with slow $V_g$ sweeping by a semiconductor parameter analyzer and a ferroelectric tester. Also, $Q_m$ *vs.* $V_g$ are measured by a quasi-static procedure with the slow $V_g$ sweeping. The $Q_m$ is metal-gate charge density induced by ferroelectric layer polarization. Dynamic time-dependent properties of FeFETs are characterized by a pulse-write-and-$V_{th}$-read (PWVR; see Fig. 12(b)) or pulse-write-and-$I_d$-read (PWIR) method. The pulse-write (PW) is done by a pulse generator. In a FeFET, while a positive or negative $V_g$ pulse is given with the $V_g$ height of $V_h$ ($> 0$) or $V_l$ ($< 0$), ferroelectric polarization responds dynamically as a function of pulse height ($V_h$ or $-V_l$) and time width ($t_w$). After ceasing PW, considerable amounts of ferroelectric polarization are retained due to ferroelectricity. This retained polarization is a function of $V_h$, $V_l$ and $t_w$. A read operation is done after PW, where $V_g$ of the transistor is swept in a narrow voltage range slowly compared to the PW time scale. A threshold voltage ($V_{\text{th}}$) is read for PWVR, and the drain current ($I_d$) is read at a fixed $V_g$ for PWIR. A threshold voltage difference ($\Delta V_{\text{th}}$) for PWVR or a current ratio between the high- and low-$I_d$-current states for PWIR can be derived for a pair of the write voltages, $V_h$ and $V_l$. $\Delta V_{\text{th}}$ by PWVR or the $I_d$ ratio by PWIR is a performance indicator of an FeFET as a nonvolatile memory transistor.

A lot of experimental results by PWVR and PWIR are available to investigate dynamic polarization switching in metal-ferroelectric-insulator-semiconductor (MFIS) type FeFETs, whose ferroelectric is $SrBi_2Ta_2O_9$ or $(Sr_{1-x}Ca_x)Bi_2Ta_2O_9$ [3−6]. The former and the latter are abbreviated as SBT and CSBT, respectively. The SBT and CSBT belong to a family of Bi-layered perovskite oxides with similar electrical properties. The coercive field of CSBT for x = 0.1 - 0.2 is 10 % larger than that of SBT. Hereafter FeFETs comprising of SBT or CSBT in the ferroelectric layers are called SBT-FeFETs. The electric properties of SBT-FeFETs are suitable for the model verification because they are reproducible with high switching endurance due to negligibly small charge injection and trapping in the FeFET gate stacks.

Despite the many experimental works of PWVR and PWIR reported in detail, there have been no correct theories based on physics for simulating them. As one of the existing models, a phenomenological Landau-Ginzburg-Devonshire (LGD) theory in ferroelectrics constructs a Gibbs free energy for the ferroelectrics where the primary energy term is expressed as a polynomial expansion form of polarization as an order parameter [7−9]. By minimizing the free energy, equilibrium or quasi-static properties are derived. The LGD theory described the transition between the paraelectric and ferroelectric phases with temperature variation across the Curie temperature [8]. It also described phase transitions between the tetragonal and orthorhombic and between the orthorhombic and rhombohedral in BaTiO$_3$, a cubic-based ferroelectric material [7]. In the Gibbs free energy, the spatial differentiation term brought non-homogeneity, by which domain patterns were described [10, 11]. The Gibbs energy for ferroelectrics can include the elastic energy term and the coupling terms of elasticity and polarization since the polarization accompanies lattice displacement of crystals. In a particular case that the ferroelectric film is divided into LGD segments, the LGD theories are called phase field models [12, 13]. The models showed coexistence of 180°- and non-180° domains under constraints of substrates [12], and polycrystalline grain growth [14].

The LGD theory has its own representation of time dependence. In the case that the electric properties without elastic terms are matters of concern, the description of time dependence is sometimes called Laudau-Khalatnikov (LK) equation [15−17]. In the phase-field models, it is called the time-dependent Ginzburg-Landau (TDGL) equation [12,18]. LK and TDGL are essentially the same. Lagrangian function, which includes a simple viscosity term of classical mechanics [19], may help us comprehensive understanding the origin of the time-dependent polarization switching. The Gibbs free energy for the ferroelectrics replaces the potential term in the Lagrangian. In deriving equations of motion, the inertia term is usually neglected for polarization reversal problems, and the LK- and TDGL-equation are obtained [20]. Overdamped (*i.e.*, strong viscosity) cases in the Lagrangian may allow us to omit the inertia term in the equation of motion. Despite the long history of theoretical studies about ferroelectric polarization switching, there are no reports which quantitatively explained the experimental PWVR and PWIR of FeFETs. In the case of LK, for example, the time-dependent form seems simplified but makes difficulties in fitting the model to the experimental. In the case of TDGL, in which ferroelectric films are separated by many segments, calculation and parameter optimization are very time-consuming because of many physical parameters included. Therefore, the conventional models are not suitable in precise describing of the experimental PWVR and PWIR of FeFETs.

This paper proposes a physical model for describing polarization variation with time in a ferroelectric film and for calculating electronic device operations of FeFETs, metal-ferroelectric-metal (MFM) capacitors, and metal-ferroelectric-insulator-metal (MFIM) capacitors. In the model, the ferroelectric film can be a polycrystalline one, and to each grain, an angle $\theta$ is assigned where $\theta$ is the angle between the film normal and the direction along the spontaneous polarization. The $\theta$ distribution in the polycrystalline film is given by experiments. In fact, for SBT-based FeFETs the angle distribution was derived by an electron backscattering diffraction (EBSD) patterns technique [21].

In each grain having $\theta$, it is also assumed that, under an electric field, a seed for polarization reversal grows along the spontaneous direction and forms a narrow column 180° reversed domain which is expanded along the sidewise direction, because 180° switching and $\pm$ 180° domain formation are subjected to occur in SBT and CSBT ferroelectrics [22-24]. For $\theta \neq 0$ cases, observations of 180°switching and 180° domain wall moving are found in [25,26].

The Kolmogorov–Avrami–Ishibashi (KAI) model [27-29] fits the physical picture mentioned above (*i.e.*, 180° domain nucleation and domain wall propagation). The KAI model provides mathematical functions of the $\theta = 0$ case describing domain expansion with time under a constant electric field. As for epitaxial thin films, many papers indicated good agreement of the KAI model with experiments [30,31], but for polycrystalline films the KAI model is asserted to be unsuitable [32]. This complicated issue concerning polycrystal films will be overcome in our present paper by introducing $\theta$ terms in grains. In order to explain wide-range log($t$) distributions of polarization switching times that a Pt/Pb(Zr,Ti)O$_3$/Pt capacitor showed, a non-KAI model called nucleation-limited-switching (NLS) model was proposed [33]. A (111)-oriented Pb(Zr,Ti)O$_3$ was used. In the NLS model, the waiting times of elementary regions are stochastic. By giving exponentially broad distribution of the waiting time, the switching phenomena across a wide log($t$) scale were described. Since the distribution function is assumed so that the numerical results reproduce the experimental ones, it is not easy to understand the physics. If we infer a possible inclusion of grains with other crystal orientation in the film, physics in the wide log($t$) characteristics might be explained by an effect of the orientation distribution. A HfO$_2$-based FeFET is reported to be consistent with the NLS [34]. The paper indicates that the grain size is ≈20 nm and the field for creating a nucleus is ≈1 MV/cm. This means that the nucleation size is already the same as the grain size, and thus, there is no space left for the created nucleus to induce a wall expansion supposed in the KAI model.

Herein we propose an extended KAI (EKAI) model which is well applicable to represent characteristics of FeFETs with polycrystalline ferroelectrics under time-dependent electric field. In the EKAI model, KAI-like pictures are adopted only inside individual grains. The EKAI describes 180° switching and ± 180° domain wall propagations in every grain separately. No wall motion propagates across adjacent grains are assumed. There are two fundamental premises of the EKAI. One is that only the electric field component along the spontaneous polarization works for the polarization switching. The other is that we pick up only the film-normal component of the switched polarization because the film-normal component is the important response to the externally applied potential. The in-plane components of the switching polarization are expected to be randomly distributed among grains. The effect of the in-plane components is thought to be weak because of the random distribution, and thus, the boundary conditions between neighboring grains are ignored in the present EKAI model. The polarization of the KAI model is an averaged quantity where the polarization is represented by the average of the volume fraction of ± 180° domain regions. The EKAI has a distinct perspective on ferroelectrics from the phase-field model in which the ferroelectrics are divided into small segments connected by strict boundary conditions according to the idea of the LGD. The EKAI has much shorter computation time than the phase-field model because of ignoring the grain boundary conditions and averaging the ferroelectric polarizations in grains. The PWVR and PWIR data show wide time range of polarization switching characteristics from 50 ns to 10 ms [3-6]. We suppose that these rather slow and wide-ranged time responses attribute to ferroelectric polycrystals consisting of grains which have broad distributions in the crystal-orientations. The response times of the paraelectric components in the ferroelectric grains and the dielectrics in the insulator are supposed to be much shorter than those of the polarization switching components. The potential formation time in the semiconductor is also supposed to be shorter. Therefore, the EKAI model in this paper assumes that the parameters except for the polarization switching in the ferroelectric grains varies instantaneously. Correctness of the EKAI model is supported by good agreements with experimental results of quasi-static $I_d - V_g$, $Q_m - V_g$, and PWVR for SBT-based FeFETs as discussed later in Sec. IV. The EKAI model in this paper assumes no free charges existing in the gate insulator and ferroelectric, thus the EKAI is not directly applicable to HfO$_2$-based FeFETs in which the ferroelectric polarization switching is always accompanied by charge injection currents [35−39].

Let us write some equations of the original KAI model, because they are necessary in the succeeding section for the EKAI derivation. In the original KAI, polarization switching nucleation is instantaneous in comparison with domain growth motion. A constant electric field and the domain wall velocity ($v_{wall}$) depending on this constant field are assumed. As shown in Fig. 1, the ferroelectric volume is constituted by domain regions. The switching polarizations in the downward and upward regions are $P_s$ and $-P_s$, respectively. $P_s$ is the spontaneous polarization. The polarization direction is parallel to the z-axis. The downward (or upward) domain regions expand with time under a positive (or negative) field, after application of step function with a constant positive field, $E_{z+}$. Consider a case that at the initial ($t < 0$) the volume is occupied fully by the upward regions, and a step function with a constant positive field, $E_{z+}$ is applied at $t = 0$. Then, the volume fraction of the downward domain ($R_{dn\|z}$) is varied with time for $t \geq 0$ under a constant $E_{z+}$ as:

$$R_{dn\|z} = c(t), \qquad (1)$$

with

$$c(t) = 1 - \exp\left(-\left(\frac{t}{t_o}\right)^n\right). \qquad (2)$$

$t_o$ is the characteristic time for the polarization switching and is a function of the constant field $E_{z+}$. The power exponent, $n$, is a parameter relating to the domain growth dimension. The volume fraction of the upward domain ($R_{up\|z}$) is:

$$R_{up\|z} = 1 - c(t). \qquad (3)$$

The switching polarization ($\bar{P}_z$) averaged over the ferroelectric volume is:

$$\bar{P}_z = R_{dn\|z} P_s + R_{up\|z}(-P_s) = P_s(2c(t) - 1). \qquad (4)$$

Similarly, in the case that at the initial ($t < 0$) the volume is occupied fully by the downward regions, and step function with a constant negative field, $E_{z-}$ is applied at $t = 0$, the volume fraction of the upward domain ($R_{up\|z}$) is for t ≥ 0 under a constant $E_{z-}$:

$$R_{up\|z} = c(t). \tag{5}$$

$t_o$ in Eq. (2) is a function of the constant field $E_{z-}$. The switching polarization ($\bar{P}_z$) averaged over the ferroelectric volume is:

$$\bar{P}_z = -P_s(2c(t) - 1). \tag{6}$$

According to Ishibashi and Takagi [29] and Ishibashi [40], in category 1, domain nucleation occurs with a fixed probability, and $n = 3$ when the wall shape is two-dimensional (*i.e.*, circular) and $n = 2$ when the wall shape is one-dimensional (*i.e.*, straight line). In category 2, there exist latent nuclei, and no new nucleation appears. In category 2, $n = 2$ when the shape is two-dimensional, and $n = 1$ when it is one dimensional. In real materials, domain nucleation may occur, and latent nuclei may also exist. Some parts are two-dimension- and others are one-dimensional-like. Hence the value $n$ is not an integer of the range $1 \le n \le 3$. When $t_o$ is small in Eq. (2), the transient time of the polarization reversal is short, when $t_o$ is large, it is long. That is, $t_o$ is a characteristic time that gives a time scale of the polarization variation. $t_o$ has a relationship with $v_{wall}$.

For the category 1,

$$t_o = \gamma_1 (v_{wall})^{(1-n)/n}. \tag{7a}$$

For the category 2,

$$t_o = \gamma_2 v_{wall}^{-1}. \tag{7b}$$

In Eq. (7), $\gamma_1$ and $\gamma_2$ are constants.

The KAI model did not show explicit mathematical forms of $t_o$. The wall velocity $v_{wall}$ as a function of the applied film-normal electric field ($E_z$) has been investigated mainly by switching current measurement of single crystals [41,42] and by piezoresponse force microscopy (PFM) of epitaxial thin films including random disorder by defects [43–45]. A consensus view in the case of films including the disorder is that $v_{wall}$ under high electric fields quickly increases in linear equation whereas it creeps up in reciprocal exponential at low fields [42,46]. In the case of ferroelectric films including defects, $v_{wall}$ under high field is expressed as $v_{wall} \propto E_z$ for $E_z \gg E_{dpin}$, where $E_{dpin}$ is a critical field over which depinning of wall motions occurs at 0 K. [46,47]. In an intermediate $E_z$ region, $v_{wall}$ is expressed using a power exponent $\tau'$ as,

$$v_{wall} \propto (E_z - E_{dpin})^{\tau'} \tag{8a}$$

and in a low-$E_z$ creep region, $v_{wall}$ is expressed at finite temperature as,

$$v_{wall} \propto \exp\left[-\frac{U'}{k_B T}\left(\frac{E_{dpin}}{E_z}\right)^\sigma\right], \tag{8b}$$

where $U'$ is a scale of energy barrier, and $\sigma$ is a power exponent originated from random disorder defects in ferroelectric films [46,47]. ($k_B$: the Boltzmann constant, and $T$: the absolute temperature).

From Eqs. (7) and (8), the $v_{wall}$ expressions are changed to $t_o$ expressions. Using a renormalized constant $\tau$, Eq. (8a) is, irrespective of the category 1 or 2,

$$t_o \propto (E_z - E_{dpin})^{-\tau} \tag{9a}$$

Similarly, irrespective of the categories, Eq. (8b) is converted to $t_o$, using a renormalized constant U, as:

$$t_o = t_{inf} \exp\left[\frac{U}{k_B T}\left(\frac{E_{dpin}}{E_z}\right)^\sigma\right], \tag{9b}$$

where $t_{inf}$ is a constant.

In the present work, experimental results of SBT-based FeFETs are used in order to demonstrate the EKAI model credibility. High endurance of the SBT FeFETs can be realized on small or at least moderate write-voltage conditions where the charge injection and trapping in the gate stack is suppressed [48,49]. Since a small or moderate write voltage bring a low electric field in the ferroelectric, we adopt Eq. (9b) rather than Eq. (9a) hereafter in this paper, i.e., Eq. (9b) will be used in calculations shown later. Further, since the separate determination of $U$ and $E_{dpin}$ is difficult, we have the following equation for $t_o$ in this paper,

$$t_o = t_{inf} \exp\left[\left(\frac{E_{act}}{|E_z|}\right)^\sigma\right], \tag{10}$$

with an activation field constant $E_{act}$ $\left(= (U/k_B T)^{-\sigma} E_{dpin}\right)$. To include $E_z$ negative cases, $|E_z|$ instead of $E_z$, is used in Eq. (10).

## II. EKAI MODEL

Let us consider a ferroelectric polycrystal film as shown in Fig. 2(a). The thickness of the film is $d_f$. The film is divided into plural grains labeled $l$ from one to $l_{max}$. All grains have the common thickness $d_f$. Grain boundaries are along the film normal, *i.e.*, the z-axis. The area of the grain $l$ is $A_l$. The direction of the spontaneous polarization of the grain $l$ is parallel to the $u_l$ axis. The angle between the $u_l$ axis and the z-axis is $\theta_l$. The range of $\theta_l$ is defined as $0 \leq \theta_l \leq 90°$. All grains have a same spontaneous polarization, $P_s$ or $-P_s$. The polarization of the direction of the $u_l$ axis is defined as $P_s$.

In Sec. IV, the EKAI model will be compared to the experimental results of MFIS FeFETs, where the ferroelectric layers consist of ferroelectric polycrystals. Experimental FeFETs with 135 nm thick ferroelectric SBT layers are available for discussion in the present EKAI model. Averaged in-plane diameters of the SBT grains are about 200nm which is larger than the ferroelectric layer thickness in the FeFETs. Therefore, we shall assume a single grain occupation along the *z*-direction or the film normal in the SBT FeFETs. Each grain is supposed to has a piler shape with a constant diameter from the film top to the bottom.

Figure 2(b) is an expanded schematic picture of the grain $l$. The grain consists of upward domain regions and downward domain regions. In the downward (or upward) domain regions, the spontaneous polarization $P_s$ (or $-P_s$) is parallel (or anti-parallel) to the $u_l$ axis. Figure 2(c) is a schematic picture focusing on a cylindrical-shape downward domain existing in the grain.

The EKAI model in this section describes polarization reversal behavior in one grain. The next section provides calculation schemes for the following specific devices including ferroelectric polycrystal: (1) MFM capacitors, (2) MFIM capacitors and (3) MFIS FeFETs.

1. Polarization dynamics under a time-independent electric field

In the grain $l$, the polarization direction is tilted from the z-axis by $\theta_l$. However, under a constant field, $E_z$, the wall-motion equations are assumed to be the same as those (Eqs. (1) - (6)) of the KAI model ($\theta_l = 0$).

Consider a case that a positive constant field $E_{z+}$ as a step function is applied at $t = 0$. At $t < 0$, the volume is occupied fully by the upward regions. Then, the volume fraction of the downward domain ($R_{dn}(l)$) is varied with time for $t \geq 0$ as:

$$R_{dn}(l) = c(t), \tag{11}$$

Remember that $c(t)$ is a function of $E_z(l)$ through $t_o$ The volume fraction of the upward domain ($R_{up}(l)$) is:

$$R_{up}(l) = 1 - c(t). \tag{12}$$

The switching polarization ($\bar{P}_z(l)$) averaged over the volume of the grain $l$ is:

$$\bar{P}_z(l) = R_{dn}(l)P_s + R_{up}(l)(-P_s) = P_s(2c(t) - 1). \tag{13}$$

Consider the opposite case that a negative constant field $E_{z-}$ as a step function is applied at $t = 0$. At $t < 0$, the volume is occupied fully by the downward regions. Then, the volume fraction of the upward domain ($R_{up}(l)$) is varied with time for $t \geq 0$ as:

$$R_{up}(l) = c(t), \tag{14}$$

The switching polarization ($\bar{P}_z(l)$) averaged over the volume of the grain $l$ is:

$$\bar{P}_z(l) = -P_s(2c(t) - 1). \tag{15}$$

2. Effect of the spontaneous polarization direction different from the z-axis

The function $c(t)$ in Eqs. (11)-(15) is $1 - \exp(-(t/t_o)^n)$, which is the same as Eq. (2). However, the $u_l$ axis is tilted from the z-axis by $\theta_l$. In the EKAI model, the characteristic time for the polarization switching $t_o$ is assumed to be

$$t_o = t_{inf} \exp\left[\left(\frac{E_{act}}{|E_z|\cos(\theta_l)}\right)^\sigma\right], \tag{16}$$

i.e., $|E_z|\cos(\theta_l)$ replaces $|E_z|$ in Eq. (10).

3. Switching polarization under time-dependent electric fields

Let us consider switching polarization evolution in the grain $l$ in the case that the electric field varies with time. At $t = t_{now}$, it is assumed that we know the volume fraction, $(R_{dn}(l))_{now}$, of the downward domains, that, $(R_{up}(l))_{now}$, of the upward domain, the switching polarization, $(\bar{P}_z(l))_{now}$, averaged over the grain volume, and the electric field, $(E_z(l))_{now}$. $(\cdots)_{now}$ means the amounts of $(\cdots)$ at $t = t_{now}$. In the following, The EKAI model provides $(R_{dn}(l))_{next}$, $(R_{up}(l))_{next}$, and $(\bar{P}_z(l))_{next}$ at $t = t_{next}$, where $t_{next} = t_{now} + \Delta t$ ($\Delta t$: a small time-increment), and $(\cdots)_{next}$ means the amounts of $(\cdots)$ at $t = t_{next}$.

The EKAI model assumes that the polarization varies during a short period $\Delta t$ as if the polarization varies under a constant field $(E_z(l))_{now}$. Note, however, that we must consider two cases, *i.e.*, the cases of $(E_z(l))_{now} > 0$ and $(E_z(l))_{now} < 0$.

Figure 3 shows an explanation for the case $(E_z(l))_{now} > 0$. Let us consider a case in grain $l$ that, at $t = t_{now}$, $R_{dn}(l)$ is $(R_{dn}(l))_{now}$ under $E_z(l) = (E_z(l))_{now}$. This status is the point A in the graph. Draw a curve of the EKAI function (Eq. (11)) at a constant field, $E_z(l) = (E_z(l))_{now}$, as the blue solid line $c(t)$ when $E_z(l) = (E_z(l))_{now}$. The line $c(t)$ has $(R_{dn}(l))_{now}$ at the point B. The time at the point B is $(t_{Kdn})_{now}$ which is obtained by solving the $c(t)$ (Eq. (2)) as

$$(t_{Kdn})_{now} = (t_o)_{now}\{-\ln[1 - (R_{dn}(l))_{now}]\}^{\frac{1}{n}} \tag{17}$$

with

$$(t_o)_{now} = t_{inf} \exp\left(\left(\frac{E_{act}}{|(E_z(l))_{now}\cos\theta_l|}\right)^\sigma\right) \tag{18}$$

We assume that the growth of $(R_{dn}(l))_{now}$ from $t = t_{now}$ to $t = t_{now} + \Delta t = t_{next}$ under $E_z(l) = (E_z(l))_{now}$ at point A is the same as the growth of $(R_{dn}(l))_{now}$ at point B under the same constant field $(E_z(l))_{now}$. The distance between the points A and B is $t_{now} - (t_{Kdn})_{now}$. The $R_{dn}(l)$ growth during $\Delta t$ at the point A can thus be calculated by a parallel-shifted function, $c(t - t_{now} + (t_{Kdn})_{now})$, and thus $(R_{dn}(l))_{next}$ in the case of $(E_z(l))_{now} > 0$ is obtained as

$$(R_{dn}(l))_{next} = 1 - \exp\left(-\left(\frac{(t_{Kdn})_{now} + \Delta t}{(t_o)_{now}}\right)^n\right) \tag{19}$$

The volume fraction of the upward domain, and the z-axis component of the switching polarization averaged over the grain $l$ are:

$$(R_{up}(l))_{next} = 1 - (R_{dn}(l))_{next} \tag{20}$$

$$(\bar{P}_z(l))_{next} = (R_{dn}(l))_{next} P_s \cos\theta_l + (R_{up}(l))_{next}(-P_s \cos\theta_l)$$

$$= \left(2\big(R_{dn}(l)\big)_{next} - 1\right) P_s \cos\theta_l \tag{21}$$

Quite similarly, we derive the following equations in the case of $(E_z(l))_{now} < 0$. The volume fractions of the upward domains and the downward domains are

$$\big(R_{up}(l)\big)_{next} = 1 - \exp\left(-\left(\frac{(t_{Kup})_{now}+\Delta t}{(t_o)_{now}}\right)^n\right) \tag{22}$$

and

$$\big(R_{dn}(l)\big)_{next} = 1 - \big(R_{up}(l)\big)_{next}, \tag{23}$$

with

$$\big(t_{Kup}\big)_{now} = (t_o)_{now}\{-\ln[1 + (R_{up}(l))_{now}]\}^{\frac{1}{n}} \tag{24}$$

The z-axis component of the switching polarization averaged over the grain $l$ is:

$$\big(\bar{P}_z(l)\big)_{next} = -\left(2\big(R_{up}(l)\big)_{next} - 1\right) P_s \cos\theta_l . \tag{25}$$

The electric field $(E_z(l))_{next}$ can be obtained using the obtained $\big(\bar{P}_z(l)\big)_{next}$ (Eq. (21) or (25)) and the electrostatic equation which depends on the device structure considered. See the next subsection. Once $(E_z(l))_{next}$ is obtained, we know now $(R_{dn}(l))_{next}$, $(R_{up}(l))_{next}$, and $(E_z(l))_{next}$. Then, the quantities of $(\cdots)_{next}$ are regarded as the quantities at $t = t_{now}$ of $(\cdots)_{now}$, we can repeat calculation.

When $\Delta t$ is infinitesimally small, Eqs. (19) and (22) can be expressed as a differential style of $R_{dn}(l)$ and $R_{up}(l)$. That is, for $(E_z(l))_{now} > 0$

$$\left(\frac{dR_{dn}(l)}{dt}\right)_{now} = \left(1 - \big(R_{dn}(l)\big)_{now}\right) \frac{n}{((t_o)_{now})^n} [(t_{Kdn})_{now}]^{n-1} \tag{26}$$

and for $(E_z(l))_{now} < 0$

$$\left(\frac{dR_{up}(l)}{dt}\right)_{now} = \left(1 - \big(R_{up}(l)\big)_{now}\right) \frac{n}{((t_o)_{now})^n} \left[(t_{Kup})_{now}\right]^{n-1} \tag{27}$$

The physical meaning of Eqs. (26) and (27) is that the differential of $c(t - t_{now} + (t_{Kdn})_{now})$ at point A in Fig. 3 equals the differential of $c(t)$ at point B.

### III. TOTAL CALCULATION SCHEME FOR DESCRIBING TIME-VARING SWITCHING POLARIZATIONS OF SPECIFIC DEVICES

Figure 4 shows schematic drawing of (a) an MFM capacitor, (b) an MFIM capacitor, and (c)

an MFIS FeFET. As stated before, the ferroelectric film consists of poly-crystal grains indexed by $l$ with $l = 1, 2, \cdots, l_{max}$. We apply a time-dependent voltage, $V_g$, to the top metal electrode of the MFM and MFIM capacitors, and to the gate metal electrode of the MFIS FeFET. We provide the ground voltage, 0 V, to the bottom metal electrode of the MFM and MFIM capacitors, and to the semiconductor substrate terminal.

Following the EKAI model in the previous section, we derived, at a moment, the z-axis component of the switching polarization $\overline{P_z}(l)$ averaged in the volume of grain $l$. We assume that the paraelectric component of the ferroelectric is isotropic. The electric displacement $\overline{D_z}(l)$ averaged in the volume of grain $l$ is represented as

$$\overline{D_z}(l) = \varepsilon_o \varepsilon_{fdi} E_z(l) + \overline{P_z}(l), \tag{28}$$

with $\varepsilon_o$ is the permittivity in vacuum, and $\varepsilon_{fdi}$ is the dielectric constant of the paraelectric component of the ferroelectric.

The metal electrodes of all the three devices are assumed to have large free carrier densities. At the metal electrode surfaces facing the ferroelectric or insulator, the electric-field penetrations are negligibly small, and thus no potential variations inside the metal can be assumed. The potential in the metal is uniformly $V_g$. Following Eq. (28) and the Gauss's law, the induced charge per unit area facing the grain $l$ is

$$\overline{Q}(l) = \overline{D_z}(l) \tag{29}$$

Because of the averaged value of $\overline{D_z}(l)$, $\overline{Q}(l)$ is also averaged in the volume of grain $l$. The total induced charges of the top metal electrode of the potential, $V_g$, is the summation of the induced charge $\overline{Q}(l)$ over all the grains. The total induced charges are defined as the charge per unit area, $Q_m$, is

$$Q_m = \frac{1}{A_{tot}} \sum_{l=1}^{l_{max}} A_l \overline{Q}(l), \tag{30}$$

where $A_l$ is the area that the grain $l$ faces the electrode, and $A_{tot} = \sum_{l=1}^{l_{max}} A_l$. This $Q_m$ is the quantities obtained experimentally in $Q_m - V_g$ measurements which are explained in (2) in APPENDIX. The switching polarization (*i.e.*, not including the paraelectric polarization) averaged over the entire ferroelectric, $P_z$, is as

$$P_z = \frac{1}{A_{tot}} \sum_{l=1}^{l_{max}} A_l \overline{P_z}(l). \tag{31}$$

Using the EKAI and electrostatic equations across MFM, MFIM, and MFIS stacks, calculation steps are expressed as follows:
1. MFM capacitor

The electrostatic equation of the MFM capacitor is

$$V_g - V_{fb} = V_f, \tag{32}$$

where $V_f$ is the voltage across the ferroelectric layer. The flat band voltage, $V_{fb}$, is the work function difference between the metal top electrode and the metal bottom electrode. Since $V_f$ does not depend on the grains, the electric field is also independent of $l$ as

$$E_z(l) = E_z = (V_g - V_{fb})/d_f. \tag{33}$$

We have Eq. (30) as

$$Q_m = \varepsilon_o \varepsilon_{fdi} E_z + \frac{1}{A_{tot}} \sum_{l=1}^{l_{max}} A_l \bar{P}_z(l). \tag{34}$$

The second term of Eq. (34) is the switching polarization averaged over the entire ferroelectric, $P_z$ (Eq.(31)).

Eqs. (21) and (25) of the EKAI model in the proceeding section can give us the $\bar{P}_z(l)$ (for $l = 1 \cdots l_{max}$) at $t = t_{next}$. We also know $V_g$ at $t = t_{next}$, because $V_g$ is externally applied. Then, $E_z$ and $Q_m$ at $t = t_{next}$ are obtained by Eq. (33) and Eq. (34), respectively. We know the behavior of the MFM capacitor at $t = t_{next}$.

$R_{dn}(l)$, $R_{up}(l)$, and $E_z(l)$ at $t = t_{next}$ are regarded as those at $t = t_{now}$. The EKAI model provides $R_{dn}(l)$ and $R_{dn}(l)$, and $\bar{P}_z(l)$ at $t = t_{next}$. Equation (33) gives us $E_z(l)$ at $t = t_{next}$. We repeat these procedures and obtain fully the time-dependent solution of the MFM capacitor.

2. MFIM capacitor

Charges appear at the top surface of the bottom electrode. We shall extend the z-axis-orientated boundaries between the adjacent grains into the insulator. We have the potential and $E_z$ relationships for each grain, i.e., we have the following for $l = 1, 2, \cdots, l_{max}$

$$V_g - V_{fb} = V_f(l) + V_i(l), \tag{35a}$$

and

$$E_z(l) = V_f(l)/d_f. \tag{35b}$$

Here, $V_f(l)$ is the voltage across the grain $l$. $V_i(l)$ is the voltage across the region in the insulator belonging to the grain $l$. At the top surface of the bottom electrode corresponding to the grain $l$, the induced charge is $-\bar{Q}(l)$, and we have for $l = 1, 2, \cdots, l_{max}$

$$V_i(l) = \bar{Q}(l)/C_i, \tag{36a}$$

with the insulator capacitance, $C_i$:

$$C_i = \frac{\varepsilon_o \varepsilon_i}{d_i}, \tag{36b}$$

and

$$\bar{Q}(l) = \varepsilon_o \varepsilon_{fdi} E_z(l) + \bar{P}_z(l). \tag{37}$$

We eliminate $E_z(l)$ from Eqs. (35), (36) and (37). Then, we derive

$$\bar{Q}(l) = \frac{C_i C_{fdi}}{C_i + C_{fdi}} \left[ V_g - V_{fb} + \frac{\bar{P}_z(l)}{C_{fdi}} \right], \tag{38}$$

with $C_{fdi}$ is the paraelectric component capacitance of the ferroelectric as

$$C_{fdi} = \frac{\varepsilon_o \varepsilon_{fdi}}{d_f}. \tag{39}$$

The EKAI model provides $R_{dn}(l)$ and $R_{dn}(l)$, and $\bar{P}_z(l)$ at $t = t_{next}$. We also know $V_g$ at $t = t_{next}$, because $V_g$ is externally applied. Equation (38) leads to $\bar{Q}(l)$ at $t = t_{next}$ for all $l$, and Eq. (37) gives us $E_z(l)$ at $t = t_{next}$, using the obtained $E_z(l)$ for all $l$. All physical quantities are derived at $t = t_{next}$. Then similarly to the case of the MFM capacitor, we can return the EKAI model calculation, and consequently we derive the time dependent numerical solution for the MFIM capacitor.

3. MFIS FET

Figure 6 summarizes the total calculation scheme for MFIS FeFETs. Regarding the semiconductor, in the MFIS stack, the acceptor concentration ($N_A$) and carrier density in the semiconductor are much smaller than the carrier density in the metal layer. Thus, the electric field penetration into the semiconductor inevitably occurs so that the semiconductor surface potential $\psi_s$ is different from the substrate potential ($V_{sub} = 0$). Imagine further a surface case where a region of an inversion state adjoins a region of an accumulation state. The surface potential cannot change abruptly on the surface. A good measure of the potential variation on the surface is the maximum depletion width, $W_m = 2\sqrt{\varepsilon_o \varepsilon_s \ln(N_A/n_i)/(e_o N_A \zeta)}$, where $\varepsilon_o$ is the permittivity in vacuum, $\varepsilon_s$ the relative permittivity of the semiconductor, $n_i$ the intrinsic carrier concentration, $\zeta = e_o/(k_B T)$, and $e_o$ the elementary charge [50]. For example, $W_m \cong 300$ nm when $N_A = 10^{16}$ cm$^{-3}$ $T = 300$ K using $\varepsilon_s = 11.9$ and $n_i = 1.45 \times 10^{10}$ cm$^{-3}$ for Si. If the size of ferroelectric grains is smaller than $W_m$, a uniform surface potential $\psi_s$ over all the grains is a good approximation (Fig. 4 (c)). Figure 5 shows a schematic drawing of a ferroelectric-insulator-semiconductor (FIS) part of the experimental MFIS FeFETs. As shown in APPENDIX, the gate stack of the experimental FeFETs is Ir/CSBT/HfO$_2$/Si. Via the crystallization annealing of the CSBT, an 2.6-nm-thick SiO$_2$ interfacial layer (IL) was formed between HfO$_2$ and Si. The HfO$_2$ and CSBT layers are 4-nm-thick and 135-nm-thick, respectively. The bilayer of the HfO$_2$ and IL forms the insulator in the MFIS. Transmission electron microscope photos confirmed a thin (about 5nm-thick) transitional layer as shown in Fig. 5 (a). The transitional layer is constituted by fine grains, ($\approx$ 5nm) whose main elements are originated from the ferroelectric CSBT. Due to the fine sizes, these grains may be non-ferroelectric, but work as a high permittivity material. The dielectric constants of this transitional layer, the HfO$_2$ layer, and the SiO$_2$ IL are typically 180, 25, and 3.9, respectively. The different values of $\bar{P}_z(l)$ (Eq. (28)) at the bottom of the ferroelectric among the grains are averaged via the in-plane pass in the transitional layer (Fig. 5 (b)). Therefore, the switching polarization at the interface between the CSBT and the insulator consisting of the bilayer of HfO$_2$ and IL can be reasonably assumed to equal the z-axis component of the polarization $\bar{P}_z(l)$ averaged over all the grains, $P_z$, that is already defined in Eq. (31). Since $P_z$ at the top surface of the insulator and $\psi_s$ at the bottom of it are uniform laterally, the electric field $E_i$ along the z-axis in the insulator and the potential $V_i$ across the insulator are also laterally uniform, which leads to a grain-independent field $E_z$ [$E_z(l) = E_z$ for all $l$] in the ferroelectric and a grain-independent potential $V_f = E_z d_f$ across the ferroelectric. The z-axis component of the electric displacement $D_z$ averaged over all the grains as well as $Q_m$ from Eqs. (28)-(30) are

$$Q_m = D_z = \varepsilon_o \varepsilon_{fdi} E_z + \frac{1}{A_{tot}} \sum_{l=1}^{l_{max}} A_l \bar{P}_z(l). \tag{40}$$

FeFETs have four gate-, source-, drain-, and substrate-terminals. Let us consider the case where the gate voltage $V_g$ is generally varied with time and other terminals are grounded ($V_s = V_d = V_{sub} = 0$). Here $V_s$, $V_d$, and $V_{sub}$ are the voltages applied on the source, drain, and substrate, respectively. The electrostatic equation across the MFIS stack is,

$$V_g - V_{fb} = V_f + V_i + \psi_s, \qquad (41)$$

where $V_{fb}$ is the flat-band voltage, *i.e.,* difference between the metal work function and semiconductor fermi level. By eliminating $E_z$ from Eqs. (40) and (41) and using the Gauss's law with noticing the capacitance definition of Eqs. (36b) and (39), we have

$$V_g - V_{fb} + \frac{1}{A_{tot}} \sum_{l=1}^{l_{max}} A_l \bar{P}_z(l) = Q_m \left( \frac{1}{C_f} + \frac{1}{C_i} \right) + \psi_s. \qquad (42)$$

If there is no interface state density between the insulator and semiconductor, the induced charge density, $Q_s$, in the semiconductor surface region has the same magnitude as $Q_m$ with the opposite polarity ($Q_m = -Q_s$). The $Q_s$ is a function of semiconductor surface potential $\psi_s$ as follows [50],

$$Q_s = \mathcal{F}(\psi_s) = \mp \frac{\sqrt{2}\varepsilon_o \varepsilon_s}{\zeta L_D} \left[ e^{-\zeta \psi_s} + \zeta \psi_s - 1 + \frac{n_{po}}{p_{po}} \left( e^{\zeta \psi_s} - \zeta \psi_s - 1 \right) \right]^{\frac{1}{2}}, \qquad (43)$$

with negative sign for $\psi_s > 0$. This equation is for *n*-channel FeFETs formed in *p*-type substrates. The equation for *p*-channel FeFETs can be rewritten appropriately. The Debye length is $L_D = \sqrt{\varepsilon_o \varepsilon_s / e_o p_{po} \zeta}$ with $p_{po} = (N_A - N_D + ((N_A - N_D)^2 + 4n_i^2)^{1/2})/2$ and $n_{po} = n_i^2/N_A$. The donor-, acceptor-, and intrinsic-carrier-densities in the semiconductor are $N_D$, $N_A$, and $n_i$, respectively, and $p_{po}$ and $n_{po}$ are the equilibrium densities of holes and electrons, respectively.

In the case that the interface states between the semiconductor and insulator are considered, the equation $Q_m = -Q_s$ is modified as

$$Q_m = -Q_s - Q_{it}, \qquad (44)$$

where $Q_{it}$ is the trapped charge at the interface per area that is expressed as

$$Q_{it} = -e_o \int N_{it}(\mathcal{E}) F_{SA}(\mathcal{E} - e_o \psi_s) d\mathcal{E}. \qquad (45)$$

Here, $\mathcal{E}$ is an electron-energy variable, $N_{it}(\mathcal{E})$ is the area density of interface-states per electron energy, and $F_{SA}$ is the Fermi-Dirac distribution function for acceptors [50].

In the case that $N_{it}(\mathcal{E})$ is approximated as a constant $D_{it}$ [V$^{-1}$cm$^{-2}$] with respect to the energy, $Q_{it}$ can be approximated as

$$Q_{it} = -e_o D_{it} \psi_s. \qquad (46)$$

For convenience of numerical root-finding calculations, we emphasize that $Q_s$ (Eq. (43)) monotonically decreases with the increase of $\psi_s$, and $Q_{it}$ (Eq. (45) or (46)) also has the same monotonical property. Equation (44) shows that $Q_m$ is a monotonically increasing function of $\psi_s$.

See Eq. (42). The right-hand side quantity is a monotonically increasing function of $\psi_s$. At $t = t_{next}$ in the previous section, we derived $\bar{P}_z(l)$ for all $l$ by either Eq. (21) or (25). $V_g$ at $t = t_{next}$ is given by an external voltage source. Thus, the left-hand side of Eq. (42) is a known constant at $t = t_{next}$. The right-hand side of Eq. (42) consists of only one variable $\psi_s$ and is a monotonically increasing function of $\psi_s$. Hence, $\psi_s$ can be uniquely determined at $t = t_{next}$. Once $\psi_s$ is determined, we derive $Q_m$ from Eqs. (43) – (46), and $E_z$ from Eq. (40) at $t = t_{next}$.

We now have $E_z$ at $t = t_{next}$, and, in the previous section, have $R_{dn}(l)$ and $R_{up}(l)$ at $t = t_{next}$ for all $l$. The set of $E_z$, $R_{dn}(l)$, and $R_{up}(l)$ at $t = t_{next}$ replaces a set of $E_z$, $R_{dn}(l)$, and $R_{up}(l)$ at $t = t_{now}$. If $E_z > 0$, $R_{dn}(l)$ and $R_{up}(l)$ at $t = t_{next}$ are obtained by Eqs. (19) and (20). If $E_z < 0$, $R_{up}(l)$ and $R_{dn}(l)$ at $t = t_{next}$ are obtained by Eqs. (22) and (23). The procedure of this section gives us $E_z$ at $t = t_{next}$. Repeating these procedures provide dynamics of the MFIS FeFETs.

The status of the polarization in ferroelectric or the electric displacement can be monitored by the drain current, $I_d$, of the sub-threshold region, which is represented as functions of the drain voltage $V_d$ and the surface potential $\psi_s$ by [50]

$$I_d = \frac{\mu k_B T}{e_o} \sqrt{\frac{\varepsilon_o \varepsilon_s p_{po} k_B T}{2}} \left(\frac{n_i}{N_A}\right)^2 \left(1 - e^{-\zeta V_d}\right) e^{\zeta \psi_s} (\zeta \psi_s)^{-1/2} . \qquad (47)$$

In this paper, all numerical results of FeFETs are obtained on the premise of $V_s = V_d = V_{sub} = 0$, whereas the practical $I_d$ measurements of n-channel FeFETs are usually on the condition of $V_d = 0.1$ V and $V_s = V_{sub} = 0$. Such a small difference in $V_d$ condition does not affect the validity of comparing results from the calculations and the measurements.

The EKAI, the formulae from Eq. (11) to Eq. (27), and the total calculation scheme for MFIS FeFETs, those from Eqs. (28) to (31) and from Eqs. (40) to (47), describe the general transient response of MFIS FeFETs under time-dependent $V_g$ conditions. The formulae also cover slowly changing phenomena. There is so-called data retention mode, in which all the quantities such as $Q_m$, $E_z$, $R_{dn}(l)$, and $R_{up}(l)$ vary very slowly, and the $\Delta t$ values in Eqs. (19) and (22) can be chosen flexibly with minimal changes of those quantities; thus, retention results of the period of days and years are obtained in a practical computation time.

## IV. CALCULATION RESULTS OF FEFETS AND COMPARISON TO THE EXPERIMENTAL

The EKAI model and the total calculation scheme of this paper is verified using experimental data of FeFETs consisting of MFIS gate stacks of Ir/CSBT/HfO$_2$/IL/Si. The experimental details were reviewed in APPENDIX. An insulator (I) in the modeled MFIS FeFET corresponds to the bilayer of IL and HfO$_2$ in a real CSBT FeFET. Capacitance of the I ($C_i$) was evaluated as $C_i$ = 0.99 µF/cm$^2$. Regarding ferroelectric (F), dielectric constant $\varepsilon_{fdi}$ of the paraelectric component is determined by the $Q_m$ – $V_g$ curves at $V_g <$ 0 or in the third quadrant of Fig. A2 because the semiconductor depletion layer does not affect at $V_g <$ 0. In the third quadrant, by taking the gradient of the curve of a $V_g$ sweep amplitude, the combined capacitance of $C_i$ and $C_{fdi}$ (Eqs. (36b) and (39)) was 0.54 µF/cm$^2$. Since $C_i$ = 0.99 µF/cm$^2$ and $d_f$ = 135 nm, $\varepsilon_{fdi}$ = 180 at room temperature was obtained and used for calculation in the EKAI. The $P_s$ direction of SBT and CSBT ferroelectrics is the *a*-axis direction of the crystal unit cell of SBT and CSBT [51, 52].

With regard to the angle $\theta_l$ and the area $A_l$, The technique of EBSD patterns could characterize the crystal orientation and grain size. The bar-graph plots in Fig. 7 show the distribution function of grains with the orientation angle $\theta$. The quantity of the vertical axis is the area of grains whose $\theta$ is in the range from $\theta$ to $\theta + \Delta\theta$ ($\Delta\theta$: the bar width). The scanning area of EBSD, $\approx 56$ μm$^2$, is not enough for statistical treatment, meaning that $\Delta V_{th}$ *versus* $\log(t_w)$ curves simulated by the EKAI model and the calculation scheme using the bare bar-graph plots were not smooth. As shown in Fig. A4, the experimental $\Delta V_{th} - \log(t_w)$ curves are very smooth. Therefore, we used a fitted smooth curve in Fig. 7 instead of the bare data. Although the smoothed curve is used in the actual calculation, the curve is digitized at every 3° to save calculation time.

When we make quasi-static calculation of $I_d - V_g$ and $Q_m - V_g$, a sinusoidal $V_g$ function is provided as $V_g = V_{fb} + V_{amp} \sin(2\pi f_s t)$. The frequency $f_s$ and the calculation time-length are typically set as 10 Hz and 0.2 s, respectively. The time step $\Delta t = 1 \times 10^{-9}$ s is commonly used in the calculation of $I_d - V_g$, $Q_m - V_g$, and PWVR. Validity of the value $\Delta t$ and accuracy of the calculated results were verified by confirming that calculations with $\Delta t = 0.5 \times 10^{-9}$ s gave the same results as those with $\Delta t = 1 \times 10^{-9}$ s.

Experimental $I_d - V_g$ curves do not suggest any electron-energy dependence of $N_{it}(\mathcal{E})$ in Eq. (45), and thus we use $D_{it}$ and Eq. (46) regarding the interface states between the insulator and semiconductor. Several series of $I_d - V_g$ simulations with various $V_{fb}$ and $D_{it}$ values were examined and compared to the experimentally obtained $I_d - V_g$. We found that $D_{it} = 4 \times 10^{12}$ V$^{-1}$cm$^{-2}$ is a good value to fit the experiment. Regarding $V_{fb}$, the suitable range was between - 0.8 V and -1.0 V, so we used $V_{fb}$ = - 0.8 V in this work.

The exponent $\sigma$ of $v_{wall}$ in Eq. (8b) and of $t_o$ in Eqs. (9b) and (10) is generally $\sigma \neq 1$ according to the PFM experiments and the analyses by phenomenological theories of random disorder potentials generated by imperfections in epitaxial films. The magnitude of $\sigma$ depends on materials and the material preparation methods [53]. Some experiments indicated $\sigma \approx 0.5$ of BaTiO$_3$ [54], 0.5 – 0.6 in PbZr$_x$Ti$_{1-x}$O$_3$ (PZT) of x = 0.2 [55], and 0.20 – 0.28 of ferroelectric organic polymer [56]. A report of a PZT showed the $\sigma$ decreased with raising defect densities intentionally [44]. Studies on epitaxial PZT films indicated $\sigma \approx 0.9$ and $\approx 1.0$ [43,46]. Regarding single crystals, those of BaTiO$_3$ and triglycine sulfate suggested the switching currents follow exponential functions without introducing decimal exponents [41,42]. In an early theoretical work on the wall propagation in a single crystal without disorders by Miller and Weinreich [57], $v_{wall}$ in a BaTiO$_3$ single crystal was expressed by an exponential form equivalent to Eq. (8b) when $\sigma$ =1. Among other subsequent discussions whether a practical $\sigma$ should equal one or not, a theoretical work by atomistic molecular dynamics simulation supported the validity of $\sigma$ =1 [58]. Zhao et al. [59] showed that polarization switching times of ferroelectric organic polymer films obeyed simply the Merz exponential law [41]. We did not use a method for intentionally introducing defects during annealing processes for ferroelectric layer crystallization of MFIS

FeFETs. Like these, there is no reason to adopt $\sigma$ of much less than 1. In the following simulation, we thus assume $\sigma=1$.

We adopt $n = 1.3$ in the function $c(t)$ that is a value reported for an epitaxial PZT film under less electric filed than 200 kV/cm [31]. It is also close to another $n$ for polymer ferroelectric films with a range from 1.0 to 1.5 [59]. The $n$ is corresponds to the dimension of the domain shape. The calculated results were insensitive to the varying $n$ from 1 to 3. We use acceptor density $N_A = 1 \times 10^{16}$ cm$^{-2}$ whose fluctuation does not affect the calculated results very much.

The remaining parameters, $E_{act}$, $t_{inf}$, and $P_s$ can be determined by curve fitting of numerical results to the experimental about the $\Delta V_{th}$ vs. $t_w$ in PWVR. The experimental results of $\Delta V_{th}$ vs. $t_w$ are found in Fig. A4 in APPENDIX. Regarding the numerical results, PWVR simulations with varying $E_{act}$, $t_{inf}$, and $P_s$ are introduced in Figs. 8 (a), (b) and (c), respectively. Every marker corresponds to a calculated point of $\Delta V_{th}$ vs. $t_w$. Using the cases of Fig. 8, we show how uniquely three parameters, $E_{act}$, $t_{inf}$, and $P_s$, are determined. We present a reference curve (the blue solid line with filled square markers in Fig. 8(a), (b), and (c), respectively), which was a numerical solution for well simulating an experimentally obtained $\Delta V_{th}$ vs. $t_w$. The reference curve was drawn using a set of parameters $E_{act} = E_o$, $t_{inf} = t_1$, and $P_s = P_o$ with $V_h = -V_l = 4$ V. The $E_o$, $t_1$, and $P_o$ are constants for the sake of explanation. The $E_{act}$ and $t_{inf}$ are deeply involved in ferroelectric polarization dynamics. The $P_s$ is an inherent parameter of the ferroelectric. As shown in Fig. 8(a), if $E_{act}$ is as large as 1.7 $E_o$, the increase of $\Delta V_{th}$ with $t_w$ is very slow, and the curve is far from the reference curve in a realistic $t_w$ range of the experimental. If $E_{act}$ is as small as 0.43 $E_o$, rapidly increasing $\Delta V_{th}$ approaches a saturated value. This is far from the log-linear styles. An appropriate $\Delta V_{th}$ can draw a log-linear curve like the reference curve at $E_{act} = E_o$. As shown in Fig. 8(b), $t_{inf}$ determines a quantity of the $\Delta V_{th}$ vs. $\log(t_w)$ curve shift in parallel along the $\log(t_w)$ axis. The polarization growth in Eqs. (21) or (25) is proportional to $P_s$, meaning that if $P_s$ is larger, the separation between a $\Delta V_{th}$ vs. $\log(t_w)$ curve written by pulses of $V_h(= -V_l)$ and the neighboring curve written by $V_h \pm 1V(= -V_l \mp 1V)$ is wider. See Fig. 8(c), where the curves are drawn in the case of $P_s = P_o$ and $1.7 P_o$, respectively. The role of $P_s$ is to adjust this separation distance to fit the experimental results. By experiencing these processes, $E_{act}$, $t_{inf}$, and $P_s$ are uniquely determined under a fixed σ. Here, "uniquely determined" means that no solutions having a new parameter set exist far from the derived parameter set.

Remember that $E_{act}$ has a meaning of an activation- or a threshold-field for domain wall motions as Eqs. (9b) and (10) indicate. The domain wall energy is affected by elastic and electric-dipole contributions in atomic scales. Theoretical works [44,57,60,61] indicated that the domain wall energy included a power exponent of $P_s$, indicating that the $E_{act}$ may also be a function of $P_s$. Since the treated temperature is only room temperature, three parameters $P_s$, $E_{act}$ and $t_{inf}$ can be searched independently to fit the experimental data. In the case that temperature is varied, $P_s$ is also changed with temperature. We must consider that $E_{act}$ is a function of $P_s$ via the domain wall energy in addition to the thermally-activation term $(k_B T)^{-1}$ (Eq. (9b)).

Figure 9 shows a fitting result of PWVR where the calculated curves compared to the experimental data at $V_h$ = 3 V, 4 V, 5 V, and 6 V. Markers represent the calculated points of $\Delta V_{th}$ vs. $t_w$ using an optimum parameter set of $E_{act}$ = 828 kV/cm, $t_{inf}$ = 8.30 × 10$^{-12}$ s, and $P_s$ = 3.0 μC/cm$^2$. Solid lines mean the experimental results of Fig. A4 introduced in APPENDIX. The calculation well simulated the experimental $\Delta V_{th}$ vs. $\log(t_w)$ throughout the wide $t_w$ range from 50ns to 0.5 ms. Other parameters for the calculation are summarized as follows: $\sigma$ = 1, $n$ = 1.3, $V_{fb}$ = − 0.8 V, $D_{it}$ = 4 × 10$^{12}$ V$^{-1}$cm$^{-2}$, $N_A$ = 1 × 10$^{16}$/cm$^3$, $d_f$ = 135 nm, $\varepsilon_{fdi}$ = 180. $d_i$ = 3.5 nm, and $\varepsilon_i$ = 3.9.

Using the same values of parameters as those solved for fitting PWVR (Fig. 9), other correlations were simulated which were quasi-static characteristics of $I_d$ – $V_g$ (Fig. 10) and $Q_m$ – $V_g$ (Fig. 11) with various $V_g$ sweep amplitude. The calculated results are drawn with thick and red-colored lines. The experimental are expressed by thin lines colored in black. In the calculations, $V_{th}$ was defined as $V_g$ at which the semiconductor surface potential $\psi_s$ was equal to 85% of $2\psi_B$, the surface strong inversion condition. (*i.e.*, $\psi_s = 2\psi_B \times 0.85$ where $\psi_B = (1/\zeta)\ln(N_A/n_i)$ [50]). Figures 10 and 11 show moderate agreements of the calculated results with the experimental. However, high curvature of $Q_m$ – $V_g$ around $V_g = 0$ V seems more emphasized in the calculated than in the experimental as shown in Fig.11. As an effort of the $Q_m$ – $V_g$ matching, for example, $D_{it}$ may be raised for decreasing the nonlinearity of $Q_m$ – $V_g$. But the attempt enhances another mismatch in $I_d$ – $V_g$ as shown in Fig. 10. Reason of the inconsistency is not clear now. Despite having some numerical mismatch remained in the curve fitting, the EKAI model and the calculation scheme qualitatively and comprehensively well simulate FeFET characteristics such as dynamic PWVR and quasi-static $I_d$ – $V_g$, and $Q_m$ – $V_g$

## V. DISCUSSION

### A. $Q_m$ versus $E_z$ correlations in PWVR of the EKAI

In the EKAI model, a $Q_m$ vs. $E_z$ correlation is calculated along a hysteresis loop as shown in Fig. 12(a). The drawing sequence is explained by corresponding $V_g$ variations with checkpoints as shown in Fig. 12(b), which are, **a', b, c, d, g, g' d', e, f, a, h, h'** and back to **a'**, repeated cyclically in this order. A positive pulse writing (PPW) is simulated passing through **a', b, c,** and **d**. A negative pulse writing (NPW) is simulated passing through **d', e, f,** and **a**. As stated in the introduction, we assume in this paper that the parameters except for the polarization switching varies instantaneously. In four segments on the $Q_m$ – $E_z$ curve, points instantaneously move from **a'** to **b**, from **c** to **d**, from **d'** to **e**, and from **f** to **a**. These curves can be written as

$$Q_m = \varepsilon_o \varepsilon_{fdi} E_z + P_{zconst}, \qquad (48)$$

where $P_{zconst}$ is a constant that each straight line has, and $P_{zconst}$ equals $P_{zb}, P_{zc}, P_{ze}$, and $P_{zf}$ for the line **a'-b**, **c-d**, **d'-e**, **f-a**, respectively.

Initially, idling write cycles are executed which consist of PPW and NPW without a $V_{th}$ reading (VR). The idling cycles have the role to make the $Q_m - E_z$ trajectory converged into the steady loop shown in Fig. 12(a). The simulation can be started from the coordinate origin at $t = 0$, where $E_z = 0$, $Q_m = 0$, and $R_{dn}(l) = R_{up}(l) = 1/2$ for all $l$. The $Q_m - E_z$ trajectory changes during the idling write cycles and soon become converged after experiencing plural cycles. Then the PWVR operation starts. After one PPW is executed, a VR is simulated passing through **d**, **g**, **g'**, **d'**. After one NPW is executed, another VR is simulated passing through **a**, **h**, **h'**, **a'**. Up to the turnaround points **g'** and **h'**, quasi-static read operations are performed after PPW and after NPW, respectively. At a reference level of $Q_m$, i.e., at points **g** and **h**, $V_{th}$ values are decided. At read, $V_g$ is swept to $V_{swp\_end}$ (i.e., $V_g$ at point **g'** and **h'**).

As discussed in III. 3, the electrostatic potential equation across the MFIS stack (Eq. (41)) is valid at any moment where $V_f = E_z d_f$ and $V_i = Q_m/C_i$. The $Q_m$ is a function of $\psi_s$ by a discussion at Eqs. (43)-(46). Consequently, at any time, $Q_m - E_z$ satisfies the rule of the following:

$$Q_m = -C_i d_f \left( E_z - (V_g - V_{fb} - \psi_s)/d_f \right) \quad (49)$$

Curves described by Eq. (49) is called a load line in this paper. Four load lines I, II, III, and IV appear in Fig. 12(a). The load lines I, II, III, and IV are the lines when $V_g$ in Eq. (49) equals $V_h$, 0, $V_l$, and $V_{swp\_end}$, respectively. Equation (49) indicates that a solution point $(E_z, Q_m)$ locates at any moment ($t_a$) on a load line having the $V_g$ value at $t = t_a$. The EKAI model decides which point on the load line is really the solution point.

The point **b**, **d**, **g**, and **a** are decided by that Eq. (48) and Eq. (49) are simultaneously satisfied. From **b** to **c** in PPW, the EKAI model describes a $P_z$ increase from $P_{zb}$ to $P_{zc}$ during the period $t_{w1}$ of $V_h$ application (Eq. (19)). Similarly, from **e** to **f** in NPW, the model describes a $P_z$ decrease from $P_{ze}$ to $P_{zf}$ during the period $t_{w2}$ of $V_l$ application (Eq.(22)).

In the period **d-g-g'-d'** and **a-h-h'-a'**, $V_{th}$ read operations are performed. $V_g$ is swept from 0 V to $V_{swp\_end}$. When $V_g = V_{swp\_end}$, g' and h' are also on the load line IV. As stated, $V_{th}$ values are decided at a reference level of $Q_m$, i.e., at points **g and h**. As also stated, $Q_m$ and $\psi_s$ have a single-valued function relationship each other. The reference level of $\psi_s$ ($\psi_{th}$) can replace the reference of $Q_m$. The $\psi_{th}$ is chosen in a sub-threshold region of the FeFETs so that $\psi_B \leq \psi_{th} \leq 2\psi_B$. The EKAI model indicates that, during a $V_g$ sweeping after NPW, there is a tendency that $P_z$ increases. If this is visible, the point **a'** separates from **a** and shifts to the $E_z$ smaller direction on the load line II. The EKAI model works during the period **d-g-g'-d'** and **a-h-h'-a'**. This period is in a data-retention stage even though no read operation is done. If the depolarization electric field is not small, $Q_m$ shifts to the $|Q_m|$ decreasing direction on the load line II. In this sense, **d'** and **a'** differ from **d** and **a**, respectively. The decrease amounts depend on the relationship among $E_{act}$, $E_z$, and $cos(\theta_l)$, as shown in Eq. (16).

Note that, in drawing the loop in Fig. 12(a), it passes through two points: the maximum induced charge ($Q_{mmax}$) and the minimum one ($Q_{mmin}$) (*i.e.*, the negative maximum one). The $Q_{mmax}$ and – $Q_{mmin}$ decide the amount of an undesirable current of the direct tunneling type or the electric-field-assisted (Fowler-Northeim, FN) tunneling type. At $Q_m = Q_{mmax} = Q_{mmin}$, |$V_{imax}$| = $Q_{mmax}/C_i$ = – $Q_{mmin}/C_i$ is defined where the $V_{imax}$ is the maximum voltage drop across the insulator. If $Q_{mmax}$= 2.0 μC/cm² and the insulator is 1.6-nm-thick SiO₂, then $V_{imax}$ = 0.93 V and the corresponding field maximum of the insulator, $E_{imax}$ = 5.8 MV/cm. Let us estimate gate leakage currents thorough the insulator. Using a high permittivity insulator or a combination of such insulators weakens the field a little, but the field is still high enough to induce the leakage currents. Investigations of the tunneling current of polysilicon/SiO₂/Si [62,63] are good references to know the impact of the charge injection for the MFIS stacks where the semiconductor is Si. The charge injection is mainly caused by the tunneling current through the IL (*i.e.*, SiO₂). We propose a significant guideline in investigating ferroelectric FETs. Imagin what happens at $Q_m = Q_{mmax}$ = 2.0 μC/cm². At this moment, the charge injection is the highest. According to Ref. 62, in the case that SiO₂ was about 1.6nm thick, the tunneling current was ≈10 A/cm² at the 5.8 MV/cm whereas it was ≈ 10⁻⁵ A/cm² in the case of about 3.2 nm thick SiO₂, at the same field. The voltage drops across the SiO₂ layer, $Q_m/C_{SiO2}$, [$C_{SiO2}$: the SiO₂ layer capacitance] at $Q_m$= 2.0 μC/cm² are 0.93 V and 1.85V for the 1.6 nm thick and 3.2nm thick SiO₂ layers, respectively. By choosing this twice thick insulator SiO₂, the tunneling current decreases ≈10⁻⁶ times, and the $V_g$ for writing increases only about 0.9 V. As this quantitative consideration indicates, there seems no way to avoid the tunneling current except for increasing the thickness for SiO₂. The same story is valid in MFIS. For preserving the nonvolatile device reliabilities, the IL SiO₂ should be moderately thick enough to avoid charge injection caused by the tunneling current. A strategy of thinning SiO₂ is logically failed.

If charge injections are not negligible, a scenario is as follows: In PPW, electrons are injected from the silicon. The electrons may mostly arrive at the metal electrode and be absorbed. However, some of them are trapped in the ferroelectric layer, the insulator, and the interface between the ferroelectric and insulator. The trapped ones near the silicon side may return to the semiconductor by tunneling back after PPW [37], but other trapped electrons remain trapped, leading to the increase of $V_{th}$ of $n$-channel FeFETs. While NPW holes are injected from the silicon. Similarly, some holes are stably trapped, leading to the decrease of $V_{th}$ of $n$-channel FeFETs. Since the number of trapped electrons after PPW and that of trapped holes after NPW are not the same, unintended $V_{th}$ shifts appear with increasing the cycle of PPW and NPW in endurance tests.

On the load lines I, II, III, and IV mentioned above, $P_z$- and $Q_m$-increase accompanies $E_z$ decrease, meaning that a capacitance $dQ_m/dE_z$ is negative. Note that, as described in the EKAI model and the total calculation scheme, the time response of the linear dielectric part is instantaneous, but $P_z$ variation needs time. Thus, while $V_g$ is swept back and forth between negative and positive voltages, the derived $Q_m$– $V_g$ and $I_d$ –$V_g$ curves inevitably draw hysteresis loops. The EKAI model does not realize the idea discussed in Ref. 64 regarding steep-slope non-hysteresis transistors during on-off operation for back-and-forth voltage sweeping.

## B. Coercive field in the EKAI model

Although the coercive fields can be read in the $Q_m$–$V_g$ hysteresis curves derived by the calculation of this model, the EKAI model of this paper does not contain the coercive field ($E_c^*$) as an explicit parameter. Let us find the relationship between the model and $E_c^*$ for a case of a single grain ferroelectric indexed by $\theta$. To take $Q_m$ vs. $E_z$ curves, a triangular shape $E_z$ as a function of time, like Fig. 13(b), whose linear slope is $K$, is supplied to an MFM capacitor. Fig. 13 (a) shows a $P_z$ vs. $E_z$ curve. During $E_z$ increasing with time, $R_{dn}$ varies from 0 to 1, like Fig. 13(c). Since $E_z = K \cdot$ time $+$ constant, $R_{dn}$ vs. time is converted to $R_{dn}$ vs. $E_z$. At a narrow region of $E_z$, $R_{dn}$ increases rapidly, as shown in Fig. 13(c). The $R_{dn}$ rapid increase corresponds to $P_z$ increase of the right-side branch of the hysteretic loop in Fig. 13(a), indicating that an $E_z$ at which the $R_{dn}$ rapid increase occurs is regarded as a coercive field. Since the $R_{dn}$ increase is rapid within a narrow range of $E_z$, a defined coercive field well approximates $E_c^*$ which is defined as the field at $Q_m = 0$ (Fig. 13(a)). The $E_z$ representing this narrow range field can be decided by the $E_z$ ($E_{cm}$) at which $dP_{dn}/dE_z$ takes a maximum (Fig. 13(d)). $E_{cm}$ is the coercive field derived analytically in the EKAI model. Exactly speaking, $E_{cm}$ approximates $E_c^*$, but $E_{cm}$ is not equal to $E_c^*$.

By starting from Eq. (26) with use of $dE_z = Kdt$ and by calculating the second derivative, $d^2R_{dn}/dE_z^2 = 0$, we have, without any approximation,

$$\exp\left[-\left(\frac{E_{act}}{E_z\cos\theta}\right)^\sigma\right] - \frac{\sigma K\cos\theta}{E_{act}}\frac{t_{inf}}{g(n)}\left(\frac{E_{act}}{E_z\cos\theta}\right)^{\sigma+1} = 0 \qquad (50)$$

with $g(n) = n[-\ln(1 - R_{dn})]^{(n-1)/n} - (n - 1)[-\ln(1 - R_{dn})]^{-1/n}$.

In Eq. (50), $K$ is the sweep slope, and $E_{act}$, $t_{inf}$ and $\sigma$ are the model simulation parameters. In $g(n)$, $n$ is the model parameters. Although $g(n)$ contains $R_{dn}$, $g(n)$ weakly depends on $R_{dn}$ variation. This means $g(n)$ can be regarded as a constant. (In the case of $n = 1$, $g(n)$ equals 1, and Eq. (50) does not contain the dimension parameter, $n$.) Hence, Eq. (50) can be solved for $E_z$, and the root of it is $E_{cm}$.

Equation (50) is a finding-root problem of $x_e = E_{act}/E_z\cos\theta$ on the condition of $g(n)=1$. Since $E_{act}$ is included in $x_e$, $E_{cm}$ is changed rapidly with $E_{act}$. Since $K$ appears only in the coefficient in Eq. (50), the coercive field $E_{cm}$ varies slowly with $K$. The solid curve is the solution of Eq. (50), where $K$ is varied with the constants of $\sigma = 1$, $E_{act} = 828$ kV/cm, and $t_{inf} = 8.30$ x $10^{-12}$ s. The filled-circle markers are the results of the full simulation of the EKAI model with MFM capacitor structures. Good agreement is confirmed between the solid line and the filled circles. The obtained $E_{cm}$ values that vary with $K$ are about 50 kV/cm. This value agrees well with the experimentally known value of SBT. In fact, the $Q_m$–$V_g$ curve of an MFM with a (100)/ (010)-oriented SBT thin film [65] showed $E_c^* = 48$ kV/cm when the swing amplitude and frequency were 225 kV/cm and 20Hz that corresponds to $K = 1.8 \times 10^4$ (kV/cm)/s. This point is added in the inset graph of Fig. 13(e) as the filled square, indicating good agreement with the solid curve by Eq. (50).

Slower sweeping cases are also solved by Eq. (50) as shown in the solid curve of Fig. 13 (e). $E_{cm}$ decreases very slowly with the decrease of log $(K)$. The figure shows $E_{cm}$= 48 kV/cm and 20 kV/cm at $K = 1.8 \times 10^4$ (kV/cm)/s and $5.5 \times 10^{-8}$ (kV/cm)/s, respectively. This means that $E_{cm}$ is 20 kV/cm when the field is swept with a cycle period of $1.6 \times 10^{10}$ s ($\approx$ 500 years) with sweeping $\pm$ 225 kV/cm. That is to say, although the model does not have a coercive field as an explicit parameter, the derived results assure nonvolatile performance.

## VI. CONCLUSION

An extended KAI (EKAI) model for describing the electrical properties of ferroelectric-gate field effect transistors was proposed. The model is physics-based and was validated via comparison with rich experimental data of metal-ferroelectric-insulator-semiconductor type FeFETs where the ferroelectrics were of SBT or CSBT Bi-layered-perovskite oxides. The model features and the results of comparison with the experimental data are summarized as follows.

(The EKAI model and the calculation scheme for FeFETs)

The orientation angle $\theta$ of each grain in the ferroelectric film and its size is assigned, where $\theta$ is the angle between the film normal and the spontaneous polarization direction. In each grain, polarization reversed domain nucleation occurs along the spontaneous polarization direction, and the domain wall expands to the lateral direction of the $P_s$ direction. In the case of large $\theta$, the time scale of transient phenomena under an electric field is much larger than that in the $\theta = 0$ case. This is the essential cause that the $\Delta V_{th}$ vs. $t_w$ of PWVR indicates log-linear relationships.

Since the ferroelectric thickness of experimental FeFETs compared to the model is 135 nm, which is comparable with the average size of grains, we assumed a single grain occupation along the z-direction in the SBT FeFETs. Each grain is supposed to has a pillar shape with a constant area from the film top to the bottom.

The electrostatic condition of the MFIS stacked structure renders a time-varying electric field in each grain in the ferroelectric film. The KAI equation about the time evolution of polarization is presented on a condition of a fixed electric field. The EKAI model represents polarization variation under the time-varying field as follows: At $t = t_{now}$, let grain $l$ have the volume fraction of the downward polarization domain $(R_{dn}(l))_{now}$ at a positive field $(E_z)_{now}$. The polarization changes from $t_{now}$ to $t_{next} = t_{now} + \Delta t$ as if the $(R_{dn}(l))_{now}$ would change under a constant $(E_z)_{now}$. In the case of negative $(E_z)_{now}$, the volume fraction variation of the downward domain $(R_{up}(l))_{now}$ is calculated similarly. At $t = t_{next}$, using obtained $(R_{dn}(l))_{next}$ or $(R_{dn}(l))_{next}$ and the external gate voltage, the electrostatic equations of MFIS stack derives the electric field $(E_z)_{next}$. By repeating this procedure from $t_{now}$ to $t_{next}$, the time-dependent behavior of FeFETs can be derived.

The characteristic time $t_o \propto \exp(\text{const.}/ E_z \cos\theta)$ in the EKAI model is a measure of switching time of respective grains. Wide distribution of $\theta$ makes the time FeFET response quite broad on the $\log(t)$ scale. Regarding the connection of the ferroelectric to the insulator and semiconductor, the polarization is averaged over the area at the bottom of the ferroelectric film. This average procedure can be accepted because semiconductors have a much smaller ability to shield polarization than metals and the transitional layer at the bottom of the ferroelectric works for averaging the polarization variation among the grains.

(Comparison with experimental data and discussion)

The parameters, $P_s$, $E_{act}$, and $t_{inf}$ can be uniquely determined in comparison with experimental data as $P_s = 3$ µC/cm², $E_{act} = 828$ kV/cm, $t_{inf} = 8.30 \times 10^{-12}$ s. Using these parameters, $I_d$ vs. $V_g$, $Q_m$ vs. $V_g$, and PWVR were calculated. The calculation results were explained consistently and entirely the corresponding experimental data.

Transient behavior can well be understood using $Q_m$ vs. $E_z$ planes. The paraelectric component in $Q_m$ ($= D_z$) of Eq. (40) instantaneously follows the electric field variation $Q_m = \varepsilon_o \varepsilon_{fdi} E_z + P_{zconst}$ (Eq. (48)). $Q_m$ always moves on the load line $Q_m = -C_i d_f (E_z - (V_g - V_{fb} - \psi_s)/d_f)$. On the PWVR measurement, $V_g$ is set to equal $V_h$ or $V_l$, at the pulse write (PW) stage and is swept in a small voltage range to find the $V_{th}$ at the $V_{th}$ read (VR) stage. At the PW stages, $P_z$ and $Q_m$ grow via the EKAI model during the pulse width $t_w$ and finally reaches the maximum or the negative maximum of $Q_m$.

In order to suppress the effect of charge trapping in the FeFET operations, there should be restrictions on the values of $Q_{mmax}$ and $-Q_{mmin}$. The guideline was proposed for avoiding the charge injection. For preserving the nonvolatile reliabilities, the IL $SiO_2$ should be moderately thick enough to avoid charge injection caused by the tunneling current.

On the load line mentioned above, the increase in $P_z$ and $Q_m$ accompanies the decrease in $E_z$, meaning that a capacitance $dP_z/dE_z$ is negative. In our model, linear dielectric response is instantaneous, but $P_z$ variation needs time, and thus, during back-and-forth sweeping of $V_g$, the derived curves of $Q_m - V_g$ and $I_d - V_g$ always draw hysteresis loops.

The EKAI model and the calculation scheme do not explicitly have a coercive electric field ($E_c^*$). When a triangular waveform field is given across the ferroelectric film for a metal-ferroelectric-metal capacitor, a simple equation (Eq. (50)) is derived, containing the field increasing rate, $K$, as well as $E_{act}$, and $t_{inf}$. The root of Eq. (50) gives an electric field $E_{cm}$ that approximates $E_c^*$. $E_{cm}$ decreases with the decrease of $K$, but even if a very slow $K$ corresponding to a time scale of more than 100 years is chosen, a sufficient $E_{cm}$ remains. This means that the EKAI model assures a non-volatile memory function which the ferroelectrics hold.

The EKAI model of this paper described polarization variation based on 180º polarization switching. The model is applicable to the materials showing 180º or nearly 180 º polarization switching such as SrBi$_2$Nb$_2$O$_9$, Bi$_4$Ti$_3$O$_{12}$, Bi$_{4-x}$Ln$_x$Ti$_3$O$_{12}$ (Ln = La, Nd, Sm), LiTaO$_3$, and LiNaO$_3$, [66-72].

## APPENDIX: SUMMARY OF EXPERIMENTAL RESULTS OF FeFETs

EKAI model was developed and verified using experimental data of FeFETs consisting of MFIS gate stacks of Ir/CSBT/HfO$_2$/IL/Si. Manufacturing details of the FeFETs were explained elsewhere [73]. The thicknesses of CSBT and HfO$_2$ were 135 nm and 4 nm, respectively. The interfacial layer (IL) was formed on the silicon substrate surface during the crystallization annealing by which the ferroelectricity appeared in the CSBT. The IL was an amorphous SiO$_2$ with the thickness of 2.6 nm according to the cross-sectional observation by transmission electron microscopy. The insulator (I) layer of the MFIS was composed of double layers of the HfO$_2$ and the IL. Since the dielectric constants of SiO$_2$ and HfO$_2$ are 3.9 and about 20, respectively, the I layer is characterized by $d_i = 3.5$ nm, $\varepsilon_i = 3.9$, and $C_i = 0.99$ µF/cm$^2$ in the calculation scheme. Gate metal length $L_m$ was 10 µm. Channel length $L$ was 8 µm as a distance between the source and drain edges, implying that the overlap lengths were 1 µm for both gate-and-drain and gate-and-source. The FeFETs of various gate widths ($W$) from 10 µm to 200 µm were on a chip ready for experimental data measurements.

To get the thin IL, the FeFETs were annealed for the CSBT crystallization in an N$_2$ main gas mixed with a small amount of O$_2$. We found the mixing ratio of the ambient gas changed the CSBT crystal orientations in FeFETs during the annealing. EBSD showed that orientation amount *vs.* the angle $\theta$ was distributed almost flat in O$_2$ annealed FeFETs [21] but not flat in N$_2$ dominant annealed FeFETs [74]. The FeFETs adopted in the present work are the ones annealed in the ambient N$_2$ dominant, where the orientation is unevenly distributed; major grains have $\theta \geq 65º$ [74] as shown in Fig. 7.

Three kinds of experimental data are compared to numerical results by the EKAI model calculations, *i.e.*,

(1) drain current ($I_d$) *versus* gate voltage ($V_g$),
(2) Induced charge $Q_m$ in the metal gate *versus* $V_g$, and
(3) $V_g$-pulse write and $V_{th}$ read, *i.e.*, PWVR.

(1) Experimental $I_d$-$V_g$

$I_d$-$V_g$ curves were obtained using a semiconductor parameter analyzer (Agilent 4156C), in which the $V_g$ sweeping was quasi-statically slow. The experimental curves of an $n$-channel FeFET are shown in Fig. A1. The source voltage $V_s$ and the substrate voltage $V_{sub}$ were 0 V, and the drain voltage $V_d = 0.1$ V. The $I_d$ was normalized by $W/L$, and $V_{th}$ was defined as the $V_g$ at $I_d = 10^{-8}$ A.

The memory window $V_w$ is defined by $V_{thr} - V_{thl}$, where the threshold voltage $V_{thr}$ ($V_{thl}$) is on the right- (left-) side on the $I_d$-$V_g$ hysteresis curve. With increasing the $V_g$ sweep amplitude, $V_w$ was increased (the inset of Fig. A1) due to increased polarization switching. By using Ir [75] or Pt [3] as a gate metal, SBT or CSBT FeFETs could have long retention and high endurance [3,75]. Since the work functions of Pt and Ir [76] are larger than the fermi level of $p$-type Si substrates, the SBT or CSBT FeFETs originally have a larger $V_{thl}$ than 1 V [3,77]. Practically, FeFETs with nearly 0 V for $(V_{thr} + V_{thl})/2$ are preferred, thus carrier concentrations are controlled by $n$-type shallow doping in the Si channel [73,78]. By adjusting the dose and energy of the dopants, $V_{thl}$ and $V_{thr}$ can be shifted to negative by more than 1 V at the cost of increase in off-state $I_d$. In the $I_d$-$V_g$ curves as shown in Fig. A1, a high off-state current of $10^{-12}$ A order is due to the channel surface doping. In the calculation scheme of this paper, this surface doping effect is treated as an adjustment of the flat-band voltage $V_{fb}$ with negative values. Presence of a large surface state density, $D_{it}$, in the channel is suggested by large SS (subthreshold swing) values of 140 mV/decade at around $I_d = 10^{-9}$ A. As shown in Fig. A1, $I_d$ increases with decreasing $V_g$ below about – 1 V. This is the gate-induced drain leakage (GIDL) current [79, 3]. Due to the overlapped large area of the gate and drain, a $p$-type inversion layer is formed at the surface of the drain beneath the gate metal, when a negative $V_g$ is applied. This $p$-inversion forms a reverse-biased $p - n$ junction in the $n$-drain region. Since the impurity concentration is heavy in the drain, the reverse-biased junction easily accompanies a tunnel current. The GIDL is closed in the substrate and related neither to the gate ferroelectric defects nor the gate leakage current between the gate and silicon. If one chooses a structure with negligible gate-and-drain overlapping, the GIDL current can be diminished. When we used a self-aligned gate structure, this current was indeed decreased [80].

(2) Experimental $Q_m$ - $V_g$

$Q_m$ vs. $V_g$ curves were obtained using a ferroelectric test system (RT6000S, Radiant Technology). The gate is connected to one terminal of the system, and the substrate, source, and drain are connected to the other terminal. The measurement time between the neighboring two points is 1.16 ms, meaning that if 200 measurement points are assigned for one loop, it takes 0.23 s to measure one loop. The experimentally obtained $Q_m$ vs. $V_g$ curves with several $V_g$ sweep amplitudes are shown in Fig. A2. Because of the MFIS structure if we fix $Q_m$ and assume no charge injection, $V_g$ is the summation of the voltages across the ferroelectric and across the insulator, and the surface potential of the silicon channel. The curves in the first quadrant of the $Q_m$ - $V_g$ coordinates include the depletion state of the semiconductor surface side, and those in the third quadrant do not include it. Hence, the curves are asymmetric for the origin of the coordinate. Despite the semiconductor surface depletion, the curve variation with changing $V_g$ in Fig. A2 is smoother than expected. This smoother phenomenon can also be explained by charge occupation and release in a rather high interface state density presented at the silicon channel surface. The gradient of the curves in the third quadrant can be utilized to decide the permittivity of the linear paraelectric component, $\varepsilon_{fdi}$, of the ferroelectric because the depletion state is not included in this side.

(3) Experimental PWVR

Two threshold voltages $V_{thn}$ and $V_{thp}$ were defined in PWVR. The $V_{thn}$ was measured after a negative single pulse $V_g$ with height ($V_l$) and width ($t_w$). The $V_{thp}$ was measured after a positive single pulse $V_g$ with height ($V_h = -V_l$) and $t_w$. The difference $\Delta V_{th}$ was evaluated by $\Delta V_{th} = V_{thn} - V_{thp}$ as a function of $V_h$ and $V_l$ at various $t_w$. The PWVR measurement was performed according to the schematic chart (Fig. A3). In FeFETs, no full saturation of ferroelectric polarization is allowed. If we would make the ferroelectric polarization fully saturated, we must apply large $V_g$ which leads to charge injection and traps unacceptable for FeFETs. As the consequence of the unsaturated polarizations, any threshold voltage $V_{th}$ of the FeFETs are not defined using a specific reference value. In other words, a single $V_{th}$ has no meaning to identify the memory state of a FeFET which can only be recognized relatively by a difference $\Delta V_{th}$ from the previous $V_{th}$. Hence, we use a pair of write voltages, ($V_h, t_w$) and ($V_l, t_w$) with $V_l = -V_h$. Idling cycles of the pairs of pulses were given plural times before performing the write and read operation. The role of the idling cycles is to obtain $V_{th}$ unaffected by the history of the previous operations. In our experiment, as shown in Fig. A3, after the twice idling a write ($V_l, t_w$) pulse is applied then $V_{thp}$ is read by sweeping $V_g$ in a narrow range from $V_{swp\_start}$ to $V_{swp\_end}$. Next, a write ($V_h, t_w$) pulse is applied then $V_{thn}$ is read by weeping $V_g$ in the same range as the $V_{thp}$ read. During the idling and the writing, $V_d = V_s = V_{sub} = 0$ V. During the read, $V_d = 0.1$ V and $V_s = V_{sub} = 0$ V. Time length of the read by sweeping $V_g$ was of the order of 1 s. $V_g$ was applied by a pulse generator Agilent 81110A, and $I_d$ was measured by Agilent 4156C. Our homemade program codes written by the LabVIEW controlled these machines. Figure A4 shows the experimental results of PWVR. The figure shows that $\Delta V_{th}$ varies linearly with $\log(t_w)$. These log-linear properties have been commonly obtained for SBT- FeFETs [81] and CSBT- FeFETs [5,6] reported previously. In writing, $t_w$ was changed from 100 ns to 1 ms, with $V_h = 3$ V, 4 V, 5 V, 6V. In reading, $V_{swp\_start} = 0$ V, $V_{swp\_end} = 1.4$V, and $V_d = 0.1$ V. Write pulses with combinations of long $t_w$ and high $V_h = -V_l = 5$ V and 6 V were not applied to avoid charge injection and trap in the FeFET.

---------------------------------------------------------------------------------------------------------------------------------------

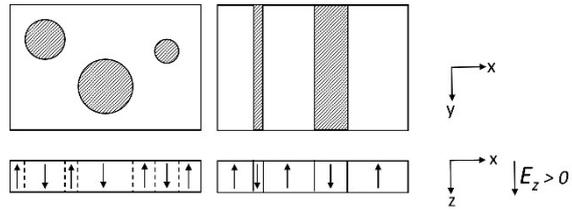

FIG. 1 KAI schematic drawing of domain nucleation and growth in a film under an electric field parallel to the z-axis ($E_z >0$). The uppers and lowers are top- and side-views, respectively. The lefts and rights represent two- and one-dimensional nucleation and growth manners, respectively. The shaded areas are on the way to expansion. An analytical function of the KAI model (Eq.(2)) describes the volume fraction increase of the downward (upward) domain with time under a constant positive (negative) $E_z$.

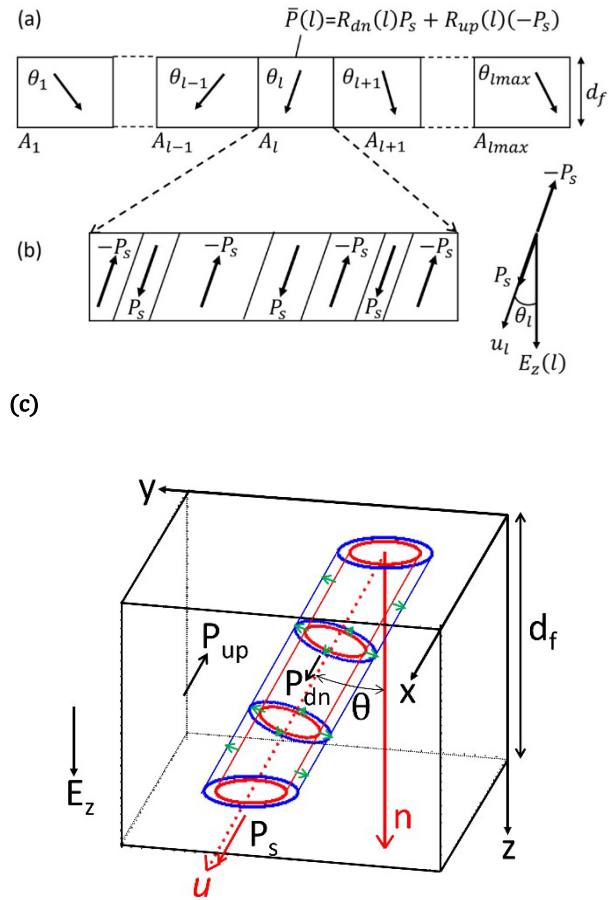

FIG. 2 (a) EKAI schematic picture of a ferroelectric film constituted by polycrystal grains. Grain $l$ has an orientation angle $\theta_l$ and area $A_l$. $\theta_l$ is the angle between the z-axis and the $u_l$ axis. The $u_l$ axis is parallel to the spontaneous polarization $P_s$. (b) Grain $l$ consists of downward ($P_s$) and upward ($-P_s$) domain regions. The $P_s$ direction in the upward (downward) domains is parallel (anti-parallel) to the $u_l$ axis. (c) A scheme of the EKAI model. A polarization nucleus is formed along the u-direction and the domain wall spreads laterally to the u-direction.

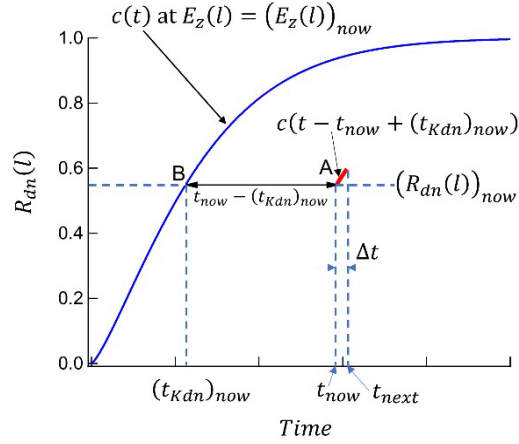

FIG. 3 Concept of the EKAI model when the electric field changes with time. In the case of $E_z$ varying with time, let us consider a case in grain $l$ that, at $t = t_{now}$, $R_{dn}(l)$ is $(R_{dn}(l))_{now}$ under $E_z(l)= (E_z(l))_{now}$. This status is the point A in the graph. Draw a curve of the EKAI function (Eq. (11)) at a constant field, $E_z(l)= (E_z)_{now}$, as the blue solid line $c(t)$ ((Eqs.(2) and (11))). The point on the curve $c(t)$ at $R_{dn}(l) = (R_{dn}(l))_{now}$ is B, and the time at the point B is $(t_{Kdn})_{now}$. We assume that the growth of $(R_{dn}(l))_{now}$ from $t = t_{now}$ to $t = t_{now} + \Delta t = t_{next}$ with $E_z(l)= (E_z(l))_{now}$ at point A is the same as the growth of $(R_{dn}(l))_{now}$ at point B under the same constant field $(E_z(l))_{now}$. The $R_{dn}(l)$ growth during $\Delta t$ at the point A can thus be calculated by a parallel-shifted function, $c(t - t_{now} + (t_{Kdn})_{now})$, and $(R_{dn}(l))_{next}$ is obtained as Eq. (19). At $t = t_{next} + \Delta t$ the same procedure is repeated under a varied new-$E_z(l)$, and we obtain $R_{dn}(l)$ at $t = t_{next} + \Delta t$. The same is performed for $R_{up}(l)$ in the case of $E_z(l) < 0$, and $(R_{up}(l))_{next}$ is derived as Eq. (22).

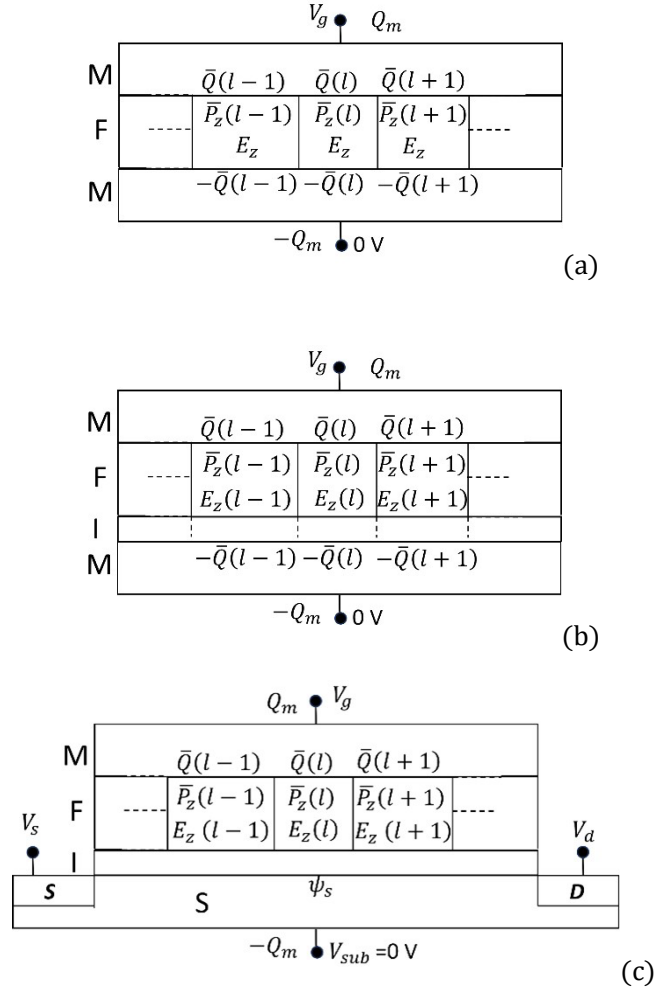

FIG. 4 Total calculation scheme for specific devices. (a) an MFM capacitor. (b) an MFIM capacitor. (c) an MFIS FeFET. The EKAI model under a varying field $E_z$ with time provides $\bar{P}_z(l)$, $R_{dn}(l)$, and $R_{up}(l)$ at $t = t_{next}$ using $R_{dn}(l)$, $R_{up}(l)$, and $E_z(l)$ at $t = t_{now}$. To repeat calculation, it is necessary to obtain $E_z(l)$ at $t = t_{next}$. Depending on the structure of the specific devices, the method of deriving $E_z(l)$ is different. For an MFM capacitor case (a), $E_z(l)$ does not depend on $l$, and $E_z$ at $t = t_{next}$ is obtained by Eq. (33). For an MFIM capacitor case (b), the induced charges of the bottom electrode of grain $l$, $-\bar{Q}(l)$ is a good approximation. $E_z(l)$ depends on $l$. $Q(l)$ at $t = t_{next}$ is derived by Eq. (38). Then $E_z(l)$ at $t = t_{next}$ is obtained by Eq. (37). For an MFIS FeFET case (c), $D_z$ at the top surface of the insulator is averaged over all the grains due to a transitional thin layer between the ferroelectric and the insulator. (See Fig. 5.) The surface potential $\psi_s$ of the semiconductor is assumed to be uniform because of insufficient acceptor density $N_A$. The uniform $D_z$ and $\psi_s$ lead to a grain-independent field $E_z$ [$E_z(l) = E_z$ for all $l$]. $\psi_s$ at $t = t_{next}$ is derived by Eq. (42). Then, $Q_m$ is obtained from Eq. (44), and $E_z$ is obtained from Eq. (40).

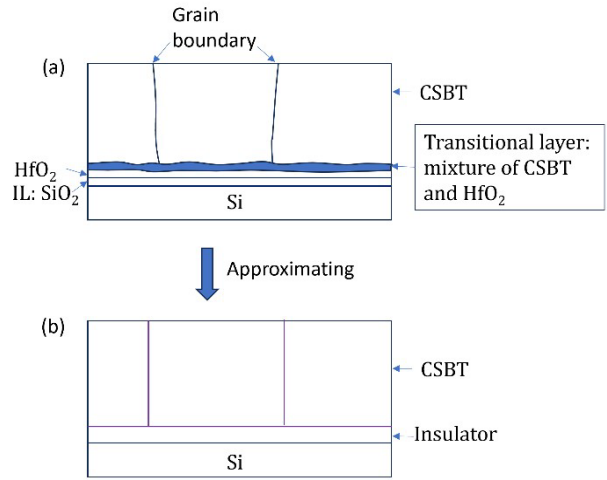

FIG. 5 (a) Schematic cross section of the experimental MFIS FeFETs. (b) The approximate structure assumed in the EKAI. In the experimental, the insulator is the bilayer of a 4 nm-thick $HfO_2$ layer and a 2.6 nm-thick IL ($SiO_2$) layer. The ferroelectric CSBT is 135 nm thick. A transitional layer exists at the interface between the CSBT and $HfO_2$ layers. A photo of a cross-section TEM confirmed this layer presence [73]. The transitional layer ($\approx$ 5 nm thick) is constituted by fine grains. The TEM photo contrast indicates that main atom elements are of the CSBT. Since a CSBT-originated high permittivity can be assumed, $D_z(l)$ is expected to be averaged of this transitional layer. Therefore, we can approximate the structure as (b) in this figure. $D_z$ is averaged over all the grains at the top surface of the insulator.

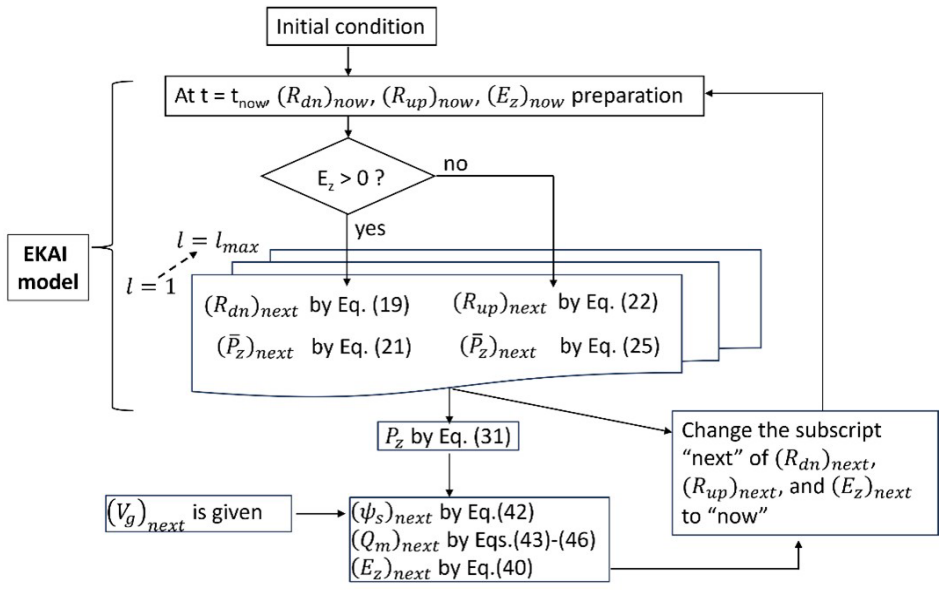

FIG. 6  Calculation scheme of MFIS-type FeFETs. The core part of the scheme is the EKAI model.

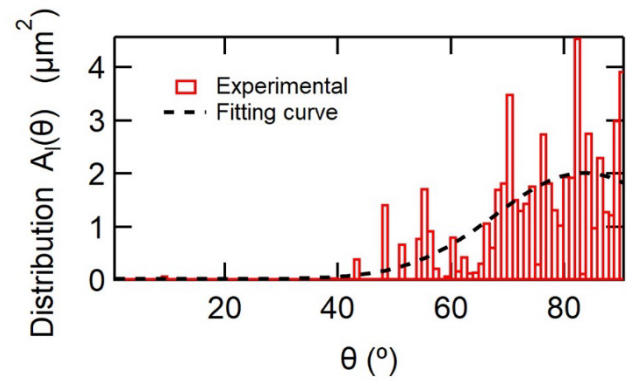

FIG. 7 Area distribution *vs.* θ of CSBT crystal orientations. The EBSD technique derived the distribution in the polycrystalline ferroelectric CSBT layer, constituting the MFIS stack in the Ir/CSBT/I/Si FeFETs. The vertical quantity of the bar graph is the total area of the grains whose angle is in the range from $\theta$ to $\theta + \Delta\theta$ ($\Delta\theta$: the bar width). The dashed line is a smoothing curve of the bar graph used in the model calculation.

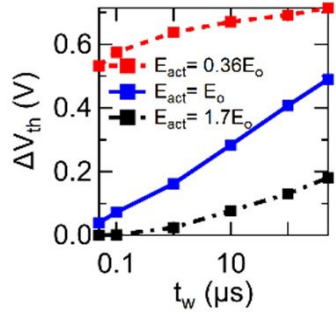

(a)

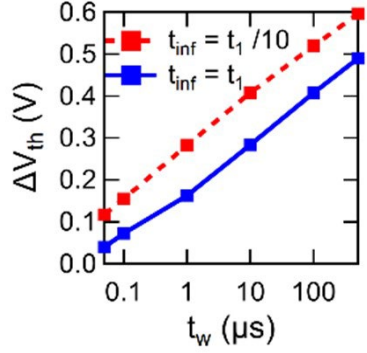

(b)

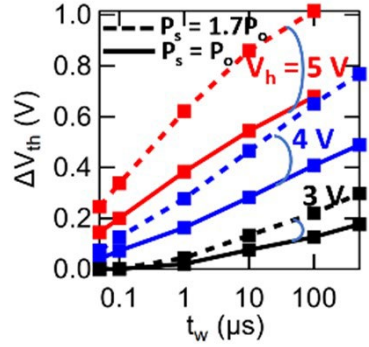

(c)

FIG. 8 Graphs explaining how the parameters, $E_{act}$, $t_{inf}$, and $P_s$, are determined when we compare the model calculations to experimental PWVR data. (a), (b) and (c) show $E_{act}$, $t_{inf}$, and $P_s$ dependences of $\Delta V_{th}$ vs. $t_w$ curves, respectively. As shown in (a), if $E_{act}$ is not optimized, the log-linear relation of $\Delta V_{th}$ vs $t_w$ cannot be derived across the $t_w$ wide range from 50 ns to 1 ms. (b) indicates that $t_{inf}$ variation brings a parallel shift of $\Delta V_{th}$ vs. $\log(t_w)$ curves. (c) indicates that the separation of $\Delta V_{th}$ vs. $t_w$ lines at $V_h = -V_l = $ 3V, 4V, and 5V is very sensitive to $P_s$. The line separation by the dashed lines is larger than the solid lines, where the former $P_s$ is 1.7 times larger than the latter. $E_o$ in (a), $t_1$ in (b), and $P_o$ in (c) are example constants for explanation convenience.

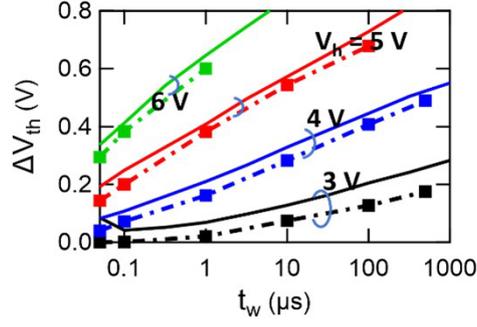

FIG. 9 Model calculation compared to the experimental data regarding PWVR. Dash-and-dot lines are the results of the calculation. Solid lines are the results of the experiment (FIG. A4). The points at the filled square markers on the $\Delta V_{th}$ vs. $t_w$ plane are the actual calculation points. Write pulse height conditions are $V_h = -V_l$ = 3, 4, 5, and 6 V. The calculation results agree fairly well with the experimental ones. In particular, the calculation reproduces the log-linear characteristics well. The significant three parameters are $E_{act}$ = 828 kV/cm, $t_{inf}$ = 8.30 × 10$^{-12}$ s, and $P_s$ = 3.0 μC/cm². Other parameters used for the calculation is summarized as $\sigma = 1, n = 1.3, V_{fb} = -0.8$ V, $d_f = 135$ nm, $\varepsilon_{fdi}= 180, d_i = 3.5$ nm, $\varepsilon_i = 3.9, D_{it} = 4 \times 10^{12}$/Vcm², $N_A= 1 \times 10^{16}$/cm³, $\varepsilon_s = 11.9$, $n_i = 1.45 \times 10^{10}$ cm$^{-3}$, and $T = 300$ K.

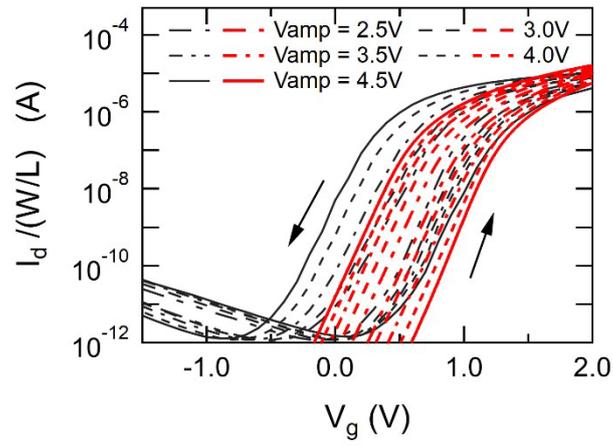

FIG. 10 Model calculation with the experiment (FIG. A1) of quasi-static $I_d$ vs. $V_g$ measurements for various $V_g$ sweep amplitude. Red lines and dark grey lines are the calculation and experiment results, respectively. We used the same parameters as those for PWVR calculation in FIG. 9.

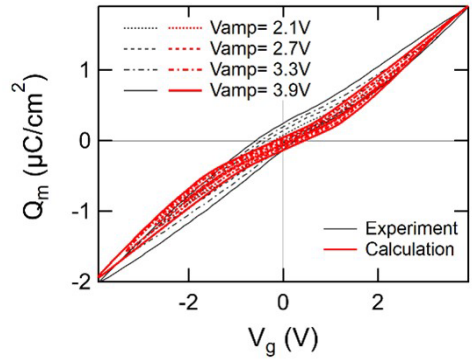

FIG. 11 Model calculation with the experiment (FIG. A2) of quasi-static $Q_m$ vs. $V_g$ measurements for various $V_g$ sweep amplitude. Thicker red lines and dark grey lines are the results of calculation and experiment, respectively. The same parameters as those for PWVR, $I_d$ vs. $V_g$ calculations in FIG. 9 and FIG. 10 were used.

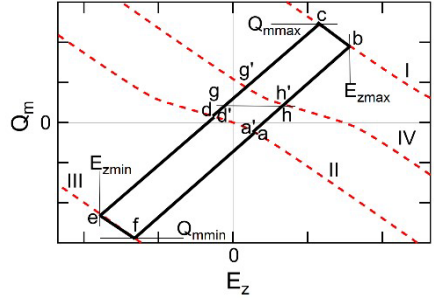

(a)

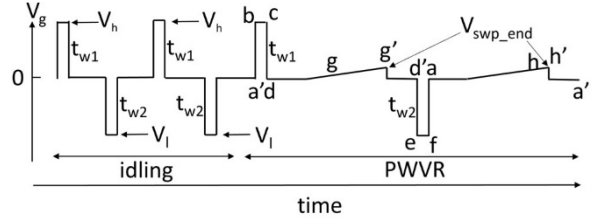

(b)

FIG. 12  (a) Solid-line loop is the trajectory on $Q_m$-$E_z$ plane during a PWVR operation, and (b) $V_g$ variation diagram for explanation. Letters from **a** to **h** with **a'**, **d'**, **g'**, and **h'** are checkpoints commonly found in (a) and (b). When a point indexed by a letter represents a state on the $V_g$ vs. time diagram in (b), the ferroelectric state takes the state on the point indexed by the same letter in the solid-line loop in (a). For a step-function-like abrupt change of $V_g$ such as **a'-b**, **c-d**, **d'-e**, or **f-a**, $Q_m$ varies instantaneously on a line whose gradient is $\varepsilon_o \varepsilon_{fdi}$. During a constant $V_g$ application like **b-c** or **e-f**, the $Q_m$-$E_z$ trajectory follows the load line I or III, $Q_m = -C_i d_f \left(E_z - (V_g - V_{fb} - \psi_s)/d_f\right)$. $R_{dn}(l)$ grows on **b-c** according to Eq. (19), and $R_{up}(l)$ grows on **e-f** according to Eq. (22) by the EKAI. The line **d-g-g'-d'** or **a-h-h'-a'** corresponds to data retention (or holding) and $V_{th}$ read stage. At the read end, $Q_m$ rises to **g'** or **h'**. At the retention stage the trajectory also exists on the load line II, $Q_m = -C_i d_f \left(E_z - (-V_{fb} - \psi_s)/d_f\right)$. Due to the presence of an electric field, the polarization may be decreased, i.e., **d** (or **a**) may move to an inside point **d'** (or **a'**) on the load line at $V_g = 0$. The decreasing extent depends on $E_{act}$, $E_z$, $\cos(\theta_l)$, and the time length between **d** and **d'** (or **a** and **a'**). $Q_m$ rising to **h'** for the read operation may also decrease the polarization.

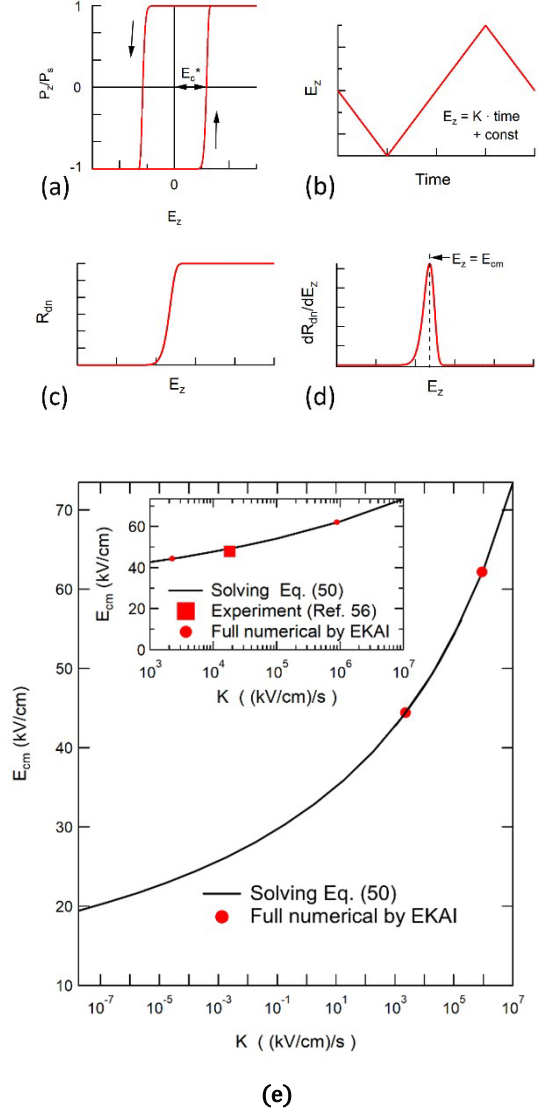

FIG. 13 Semi-analytical discussion of a quantity equivalent to a coercive field in the EKAI model. For an MFM with a single grain ferroelectric, schematic curves of (a) $P_z/P_s$ vs. $E_z$, (b) triangular waveform $E_z(t)$, (c) $R_{dn}$ vs. $E_z$, and (d) $dR_{dn}/dE_z$ vs. $E_z$. The slope, $dE_z/dt$, is $\pm K$. Since the field ($E_{cm}$) at which $dR_{dn}/dE_z$ takes a maximum is very close to the coercive field ($E_c^*$) defined as the field at $Q_m = 0$, $E_{cm}$ can be regarded as a coercive field. An analytical equation, Eq. (50), is derived, the root of which is $E_{cm}$. The solid lines in (e) and the inset of (e) are the root curve of Eq. (50) as a function of $K$. The inset in (e) is the graph in the range of $10^3 < K < 10^7$. For usual $Q_m$ – $E_z$ measurement, $K$ is within this range. The filled circles are the $E_c^*$ values obtained by the full numerical calculation of the EKAI and scheme (FIG. 6), and the filled square mark in the inset is the experimental result of SBT [63]. The solid line in (e) indicates that $E_{cm}$ is maintained as about 20 kV/cm under an ultraslow rate $K = 5.5 \times 10^{-8}$ (kV/cm)/s, corresponding that $E_z$ of $\pm$ 225 kV/cm sweeping is executed by spending $8 \times 10^9$ s.

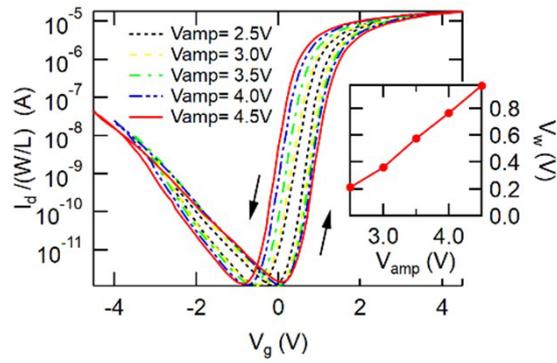

(FIG14)

FIG. A1 Experimental $I_d$ vs. $V_g$ characteristics of an Ir/CSBT/I/Si FeFET at various $V_{amp}$. $V_g$ was swept between $\pm V_{amp}$ with $V_d = 0.1$ V. Insulator I is the bilayer of HfO$_2$ layer and SiO$_2$-like IL layer. The thickness is 135 nm, 5 nm, and 2.6 nm for CSBT, HfO$_2$, and IL, respectively. The $I_d$ that is normalized to $W/L$ is shown in the figure. $L = 10$ μm and $W = 80$ μm. The small window in the figure shows in the figure $V_w$ vs. $V_{amp}$, where $V_w = V_{thr} - V_{thl}$, and $V_{thr}$ ($V_{thl}$) is the $V_g$ value at $I_d/(W/L) = 10^{-8}$ A of the right- (left-) side branch of the hysteresis curves.

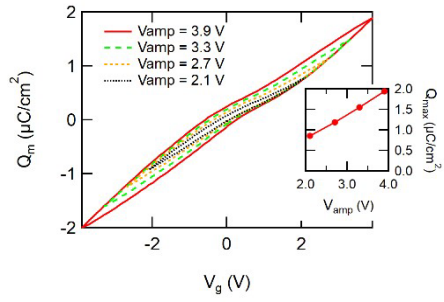

(FIG 15)

FIG. A2 Experimental $Q_m$ vs. $V_g$ characteristics for Ir/CSBT/I/Si FeFET for various $V_{amp}$. $V_g$ was swept between $\pm V_{amp}$. The stack material and thickness of each layer are the same as those of the FeFET in Fig. A1. $Q_m = Q_{max}$ at $V_g = V_{amp}$, and $Q_m = Q_{min}$ at $V_g = -V_{amp}$. $Q_{max} \cong -Q_{min}$. $Q_{max}$ vs. $V_{amp}$ is shown in the small window on the right side.

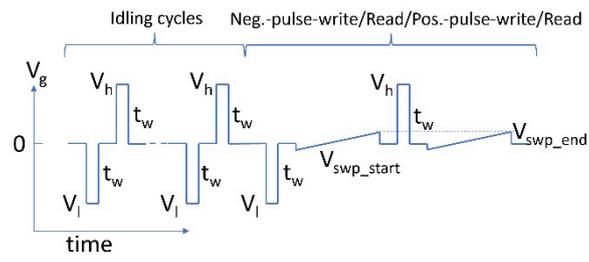

(FIG 16)

FIG. A3  Diagram of $V_g$ vs. time used for the experimental PWVR measurement. For a set of $V_h$, $V_l$, and $t_w$, after plural times repeating the idling cycle by the ($V_l$, $t_w$) and ($V_h$, $t_w$) pulses, the procedure of $V_l$ (<0) pulse write, $V_{th}$ read, $V_h$ pulse write, and $V_{th}$ read was performed.

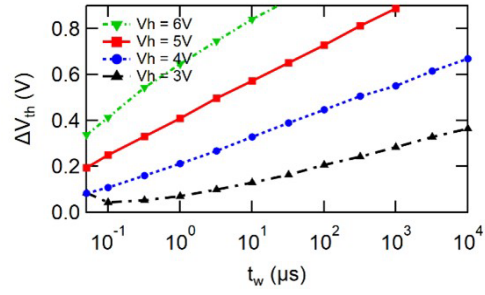

(FIG 17)

FIG. A4  $\Delta V_{th}$ vs. $t_w$ for the experimental Ir/CSBT/I/Si FeFET. The stack material and thickness of each layer are the same as those of the FeFET in Figs. A1 and A2. $t_w$ was changed from 50 ns to 10 ms, with $V_h = -V_l = 3$ V, 4 V, 5 V, 6 V for writing. $V_{swp\_start} = 0$ V, $V_{swp\_end} = 1.4$V, and $V_d = 0.1$V for reading. The markers in the graph are to show measured results. The lines connecting the markers show log-linear relationships of $\Delta V_{th}$ and $t_w$. In particular, the line for $V_h = 4$ V indicates the log-linear line across more than five orders of $t_w$.